\documentclass[traditabstract, twocolumns]{aa}

\usepackage{graphicx}
\usepackage{epstopdf}
\usepackage{deluxetable}
\usepackage{times,epsfig}
\usepackage{mathrsfs}  
\usepackage{natbib}
\usepackage{amssymb}
\bibpunct{(}{)}{;}{a}{}{,} 
\usepackage[english]{babel}
\usepackage{longtable}
\usepackage{caption}
\usepackage{url}
\usepackage{hyperref}
\usepackage[utf8]{inputenc}
\usepackage{subcaption}
\usepackage{color}

\authorrunning{A. Nyholm et al.}
\titlerunning{The bumpy light curve of supernova iPTF13z}

\begin{document}

\title{The bumpy light curve of Type IIn\\ supernova iPTF13z over 3 years\thanks{Full Tables~\ref{p48r13z_early},~\ref{13zp60P48} are only available at the CDS via anonymous ftp to \url{cdsarc.u-strasbg.fr} (\url{130.79.128.5}) or via \url{http://cdsweb.u-strasbg.fr/cgi-bin/qcat?J/A+A/605/A6}}}

\author{A. Nyholm \inst{1}
\and J. Sollerman \inst{1}
\and F. Taddia \inst{1}
\and C. Fremling \inst{1}
\and T. J. Moriya \inst{2}
\and E. O. Ofek \inst{3}
\and A. Gal-Yam \inst{3}
\and A. De Cia \inst{4}
\and R. Roy \inst{1}
\and M. M. Kasliwal \inst{5}
\and Y. Cao \inst{5}
\and P. E. Nugent \inst{6,7}
\and F. J. Masci \inst{8}}

\institute{Department of Astronomy and the Oskar Klein Centre, Stockholm University, AlbaNova, 10691 Stockholm, Sweden\\(\email{anders.nyholm@astro.su.se})
\and Division of Theoretical Astronomy, National Astronomical Observatory of Japan, 2-21-1 Osawa, Mitaka, Tokyo 181-8588, Japan
\and Benoziyo Center for Astrophysics and the Helen Kimmel Center for Planetary Science, Weizmann Institute of Science, 76100 Rehovot, Israel
\and European Southern Observatory, Karl-Schwarzschild-Str. 2, D-85748 Garching bei M\"{u}nchen, Germany
\and Astronomy Department, California Institute of Technology, Pasadena, CA 91125, USA
\and Department of Astronomy, University of California, Berkeley, CA 94720-3411, USA
\and Lawrence Berkeley National Laboratory, 1 Cyclotron Road, MS 50B-4206, Berkeley, CA 94720, USA
\and Infrared Processing and Analysis Center, California Institute of Technology, MS 100-22, Pasadena, CA 91125, USA}

\date{Received / Accepted}

\abstract{A core-collapse (CC) supernova (SN) of Type IIn is dominated by the interaction of SN ejecta with the circumstellar medium (CSM). Some SNe IIn (e.g. SN 2006jd) have episodes of re-brightening (''bumps'') in their light curves. We present iPTF13z, a Type IIn SN discovered on 2013 February 1 by the intermediate Palomar Transient Factory (iPTF). This SN showed at least five bumps in its declining light curve between 130 and 750 days after discovery. We analyse this peculiar behaviour and try to infer the properties of the CSM, of the SN explosion, and the nature of the progenitor star. We obtained multi-band optical photometry for over 1000 days after discovery with the P48 and P60 telescopes at Palomar Observatory. We obtained  low-resolution optical spectra during the same period. We did an archival search for progenitor outbursts. We analyse the photometry and the spectra, and compare iPTF13z to other SNe IIn. In particular we derive absolute magnitudes, colours, a pseudo-bolometric light curve, and the velocities of the different components of the spectral lines. A simple analytical model is used to estimate the properties of the CSM. iPTF13z had a light curve peaking at $M_r \lesssim -18.3$ mag. The five bumps during its decline phase had amplitudes ranging from 0.4 to 0.9 mag and durations between 20 and 120 days. The most prominent bumps appeared in all the different optical bands, when covered. The spectra of this SN showed typical SN~IIn characteristics, with emission lines of H$\alpha$ (with broad component FWHM $\sim 10^{3}-10^{4} ~{\rm km ~s^{-1}}$ and narrow component FWHM ${\rm \sim 10^2 ~km ~s^{-1}}$) and \ion{He}{I}, but also with \ion{Fe}{II}, \ion{Ca}{II}, \ion{Na}{i}~D and H$\beta$ P Cygni profiles (with velocities of $\sim 10^{3}$ ${\rm km ~s^{-1}}$). A pre-explosion outburst was identified lasting $\gtrsim 50$ days, with $M_r \approx -15$ mag around 210 days before discovery. Large, variable progenitor mass-loss rates ($\gtrsim$ 0.01$ ~M_{\sun} ~yr^{-1}$) and CSM densities ($\gtrsim$ 10$^{-16}$ g~cm$^{-3}$) are derived. The SN was hosted by a metal-poor dwarf galaxy at redshift $z = 0.0328$. We suggest that the light curve bumps of iPTF13z arose from SN ejecta interacting with denser regions in the CSM, possibly produced by the eruptions of a luminous blue variable progenitor star.}


\maketitle

\section{Introduction\label{sec:intro}}
The initial mass and mass-loss history of a giant star are decisive factors  for which kind of supernova explosion will end its existence. A supernova (SN) with an optical spectrum characterised by Balmer emission lines with narrow central components is classified as a SN Type IIn \citep{schlegel90, filippenko97}. Since SNe IIn are dominated by the interaction of SN ejecta with the circumstellar medium (CSM), these SNe give the observer the opportunity to probe the late mass-loss histories of their progenitor stars. These SNe are intrinsically rare. The volume limited sample by \cite{li2011} showed that 9\% of all SNe Type II are Type IIn.

The light curves of SNe IIn display great variety in peak magnitude, decline rate, and light curve evolution. From the SN~IIn samples presented by \citet{kiewe12} and \citet{taddia13}, we see that $-19 \lesssim M_r \lesssim -17$ is a characteristic range of peak absolute magnitudes for this SN subtype. For a volume-limited sample of 21 SNe IIn, \citet*[Table 2]{richardson14} gives the mean $B$-band peak magnitude as $-18.62$, with a 1.48 magnitude standard deviation of the mean. This spread of peak magnitudes is the largest for any SN subtype in their study.

After exploding, SNe IIn often reach peak brightness quickly. The 15 SNe IIn in the rise-time analysis by \citet{ofek14rise} have a mean rise time of $17\pm12$ days. These short rise times means that  SNe IIn are often discovered after peak brightness, making the time of explosion difficult to determine. After peak brightness, SNe IIn decline at a wide range of rates. This diversity is used in the literature to distinguish between different SNe IIn subtypes according to how fast they fade in visible light (e.g. \citealt[table 15;]{habergham14} and \citealt{taddia13,taddia15}). Slowly evolving, long-lasting SNe IIn are often compared to SN 1988Z \citep{aretxaga99}, which decayed at $\approx 0.2\rm ~mag ~(100 ~days)^{-1}$ at epochs $> 1$ year. A common example of rapidly developing SNe IIn is SN 1998S \citep{liu00, fassia00}, which faded at $\approx 4\rm ~mag ~(100 ~days)^{-1}$ during the first 100 days after peak. Another subtype of SNe IIn is represented by SN 1994W \citep{sollerman98}, which displayed a light curve with an initial plateau in brightness for $\sim 100$ days, followed by a sharp decline (initially $> 10\rm ~mag ~(100 ~days)^{-1}$). 

The decline after peak brightness of some SNe IIn is interrupted by episodes of re-brightening. Following e.g. \citet{margutti14}, \citet{graham14}, and \citet{martin15}, we will refer to such re-brightenings followed by fading as `bumps' in the light curve. The bumps can differ in amplitude and duration, from the short ($< 5$ days), early ($\sim~10$ days after max) bump of SN 2005la \citep[SN~IIn/Ibn hybrid, ][]{pastorello08} to the long ($\approx 500$ days), late ($\approx 500$ days after max) bump of SN 2006jd \citep{stritzinger12}. The peculiar SN 2009ip, which will be discussed in greater detail later, also had bumps in its light curve \citep[e.g.][]{margutti14, martin15}. In this paper we  present a SN~IIn characterised by multiple bumps in its light curve.

The search for progenitor stars of core-collapse (CC) SNe using archival images of sites where SNe eventually exploded led to the identification of tens of good candidates \citep{smartt15}. The rarity of SNe Type IIn has prevented this technique from being used  in most cases. The strongest Type IIn progenitor case is for SN 2005gl, where \citet{galyam07} and \citet{galyam09} used Hubble Space Telescope (HST) archival images to identify the progenitor as a luminous blue variable (LBV) star with $\rm M_V \approx -10$ mag and mass $> 50\rm ~ M_{\sun}$.

An alternative path to SN~IIn progenitor identification is to compare the mass-loss rates inferred from the CSM interaction with those known for giant stars. Such studies can support a claim of LBV stars being SN~IIn progenitors, as the wind speeds inferred from SN~IIn CSM ($\sim 100 \rm ~km ~s^{-1}$) as well as the mass-loss rates (e.g. $\sim 10^{-3} \rm ~M_{\sun} ~yr^{-1}$) conform with LBVs \citep[see e.g.][]{kiewe12,taddia13}. The need for a significant CSM surrounding the progenitor star also conforms with the episodic  mass-loss outbursts displayed by such LBVs as $\eta$ Car, like its eruptions in 1843 and 1890 \citep[e.g.][]{smith11_lbv}.

Outbursts of SNe IIn progenitors $\sim 1$ year before explosion are reported and interpreted to be common by \citet{ofek14outb}. See also \citet{bilinski15}. Among the events showing pre-explosion outbursts, SN 2009ip stands out (\citealp{pastorello13,fraser13a, ofek13b, graham14, margutti14, martin15}). SN 2009ip reached a peak magnitude of $M_R \approx -15$ in 2009 \citep{pastorello13}, indicating that it could be a SN impostor \citep{vandyk00}. In 2012, two events took place, referred to as 2012A (reaching $M_R \approx -15$) and 2012B (reaching $M_R \approx -18$). No conclusive evidence has been presented that indicates whether a destructive SN or eruptions of a still intact star were seen \citep{fraser15, smith16}. Possible analogues to SN 2009ip are SN 2010mc \citep{ofek13a}, LSQ13zm \citep{tartaglia16} and SN 2015bh \citep{thone17, ofek16, eliasrosa16}.

The environments of SNe IIn can yield important information about their progenitor channels, as demonstrated by \citet{habergham14}. \citet{taddia15} demonstrate that the metallicity at SN~IIn sites is higher than at sites of SN impostors, implying that the LBV channel is only one of the possible progenitor channels for SNe IIn.

In this paper, we introduce a SN Type IIn discovered and followed by the intermediate Palomar Transient Factory (iPTF). This transient, iPTF13z, was discovered on 2013 February~1 in a dwarf galaxy. The discovery happened after light curve peak and we have been able to follow it photometrically for over 1000 days, along with optical spectroscopic coverage. Whereas the spectra of iPTF13z conform well with common SN~IIn spectra, the light curve is remarkable. iPTF13z displays five pronounced bumps in its $r$-band light curve, with amplitudes between 0.4 and 0.9 mag. From pre-discovery images, we were also able to detect a strong candidate for a pre-explosion outburst of the progenitor at around 210 days before discovery.

In Sect. \ref{sec:obs} of this paper, we present the details of the discovery and the optical photometry and spectroscopy of the SN. Section \ref{sec:analysis} gives distance and extinction towards the SN, analyses the photometry and spectroscopy, presents our archival search for pre-explosion outbursts, and presents a study of the host galaxy. We also analyse the photometry with emphasis on the light curve bumps, make comparisons to other SN~IIn light curves, and give a simple analytical model of the CSM interaction. In Sect. \ref{sec:discussprog} we discuss progenitor star candidates and point out possible future investigations.

\section{Observations and data reduction\label{sec:obs}}

\begin{figure*}
   \centering$
    \includegraphics[width=16cm,angle=0]{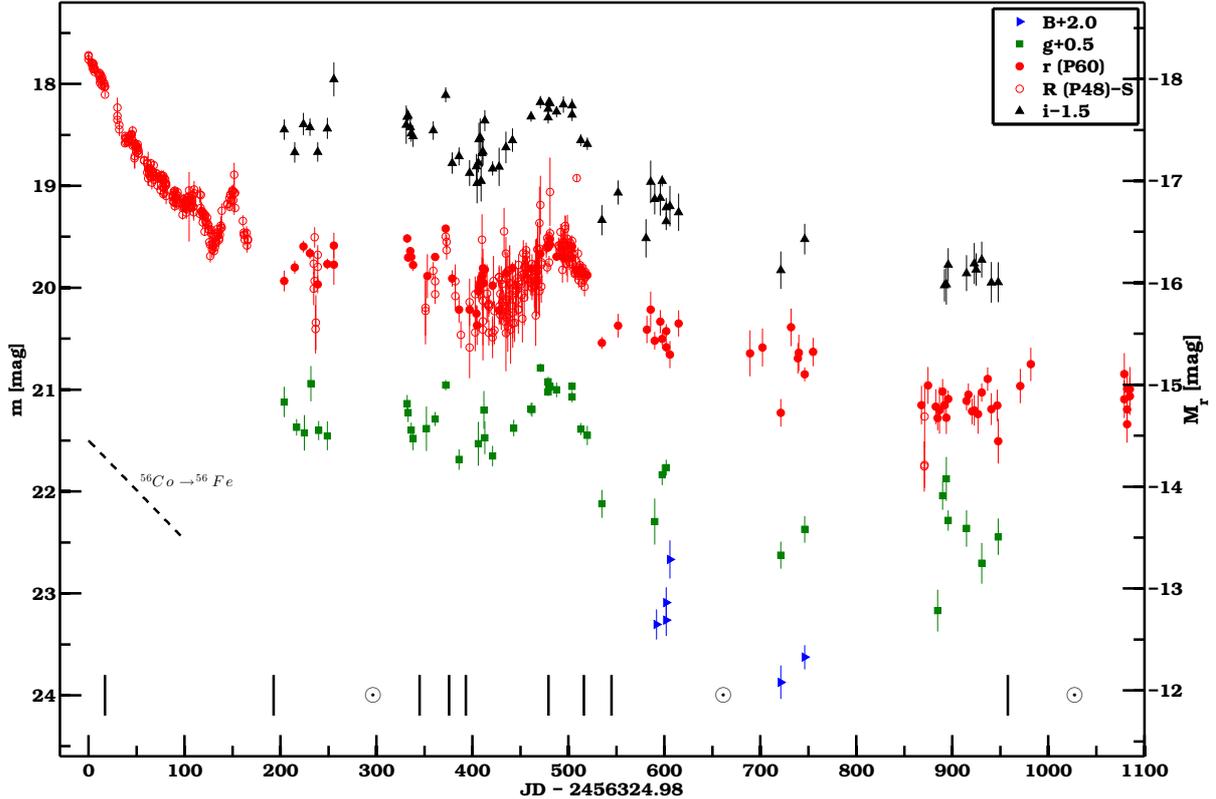}$
    \caption{Light curves of supernova iPTF13z in the $R$ band from the P48 telescope and in SDSS $g$, $r$, $i$, and Johnson $B$ bands from the P60 telescope. The $B$, $g$, and $i$ magnitudes have been offset for clarity, as indicated in the legend. The apparent magnitudes are shown without correction for extinction. Times for conjunction of iPTF13z with the Sun are marked with \sun\ symbols. The rate of the $^{56}$Co $\rightarrow~^{56}$Fe decay has been drawn for comparison. The black ticks at the bottom indicate epochs of spectroscopy. The $S$ stands for the 0.26 mag correction of the P48 photometry onto the SDSS system. The absolute magnitude $M_r$ is given for the $r$ band and takes MW extinction into account.}
    \label{fig:lc13z}
\end{figure*}

\begin{figure}[h!]
   \centering
    \includegraphics[width=9cm]{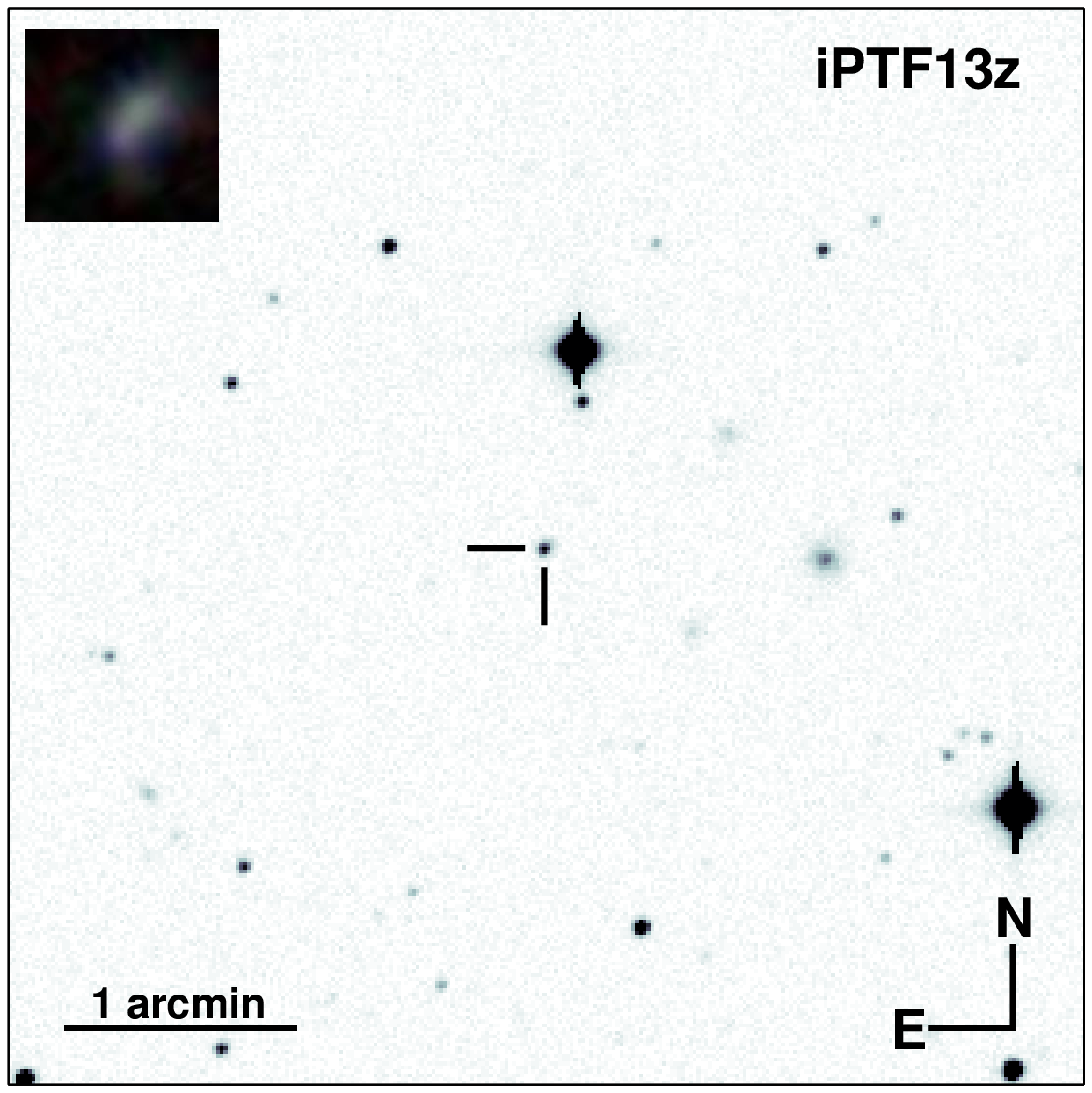}
    \caption{Finding chart for iPTF13z based on the discovery image (Sect. \ref{sec:disc}) from the P48 telescope, taken 2013 February 1.48 UT. Two black bars point at the SN. The inset image (upper left) is a $10\arcsec \times 10\arcsec$ colour composite from SDSS DR12 showing the host galaxy with the same orientation as the main image.}
    \label{fig:finder13z}
\end{figure}

\subsection{Discovery} 
\label{sec:disc}

The transient iPTF13z was first detected in an image taken in the Mould $R$ band \citep{law09} on 2013 February 1.48 UT\footnote{UT dates will be used throughout this paper.} with the 1.2 m Samuel Oschin telescope \citep[P48;][]{rahmer08} on Palomar Mountain (California, USA). The discovery was made by the iPTF \citep{kulkarni13, law09, rau09}. The image processing pipeline is described by \citet{laher14} while the photometric calibration and system is discussed by \citet{ofek12}. Point spread
function (PSF) photometry in the discovery image gave $R_{Mould}~=~17.98$ as the apparent magnitude of the transient. It was found at $\alpha = ~\rm 16^h$$\rm 02^m$$0\fs12$, $\delta =$ $21\degr 14\arcmin 41\farcs4$ (J2000.0). This is $1\arcsec$ south of the catalogued position \citep{mcconnachie09} of the irregular galaxy SDSS J160200.05+211442.3.

The latest PTF non-detection of iPTF13z is from an image with limiting magnitude $R_{Mould} = 20.56$ taken on 2012~July~28.34 UT with the P48 telescope. The (i)PTF coverage gap is thus 188 days; during this period   conjunction with the Sun took place on 2012 November 24. We searched public telescope archives for images of the iPTF13z field taken during the period 2012 July 29 -- 2013 January 31 to shrink this coverage gap. In the Catalina Real-time Transient Survey (CRTS; e.g. \citealp{drake09}) Data Release 3 we found images of the field from 2012~September 17 and 2012~September~25, taken without a filter with the 0.68 m Schmidt telescope at Mount Bigelow (Arizona, USA). With limiting magnitudes in the $18 - 19.5$ mag range, the CRTS images barely show the host galaxy and nothing is visible at the position of iPTF13z. The 2012~September~25 images from CRTS brings the coverage gap (from last non-detection to discovery) down to 129 days. This gives a weak constraint on the time of explosion for this transient.

In this paper, the age of iPTF13z will be given with respect to the time of first detection ($\rm JD~ 2456324.98$). All times are given in the observer frame.

\subsection{Photometry\label{sec:phot}}
Optical imaging of iPTF13z was done with two telescopes on Palomar Mountain. The P48 was used with the 12K$\times$8K CCD mosaic camera and a Mould $R$-band filter. The 1.52 m telescope (known as P60) was used with a 2048$\times$2048-pixel CCD camera \citep{cenko06} and filters in the SDSS (Sloan Digital Sky Survey) $g$, $r$, $i$ \citep{fukugita96}, and Johnson $B$ \citep{bessell90} bands.

Photometry of iPTF13z from both telescopes was done using the pipeline known as FPipe \citep{fremling16}. This pipeline performs sky and host subtraction and uses a PSF fit to the SN in order to perform relative photometry. The resulting photometry is given in the AB magnitude system. A selection of 29 SDSS stars (Table~\ref{refstars13z}) in the field surrounding iPTF13z was used as reference stars for the P48 photometry. No colour terms were applied. The host subtraction for P48 $R-$band images was done using a template based on 65 co-added images taken with P48 between 2009 December 21 and 2010 September 24. This time range was selected to include the deepest P48 images before detection (Sect. \ref{sec:outburst}). For the P60 photometry, host subtraction in each filter was done using a mosaic composed of 12 co-added SDSS frames \citep{ahn12}  taken in 2004 April and 2004 June as templates. See Fig. \ref{fig:lc13z} for the iPTF13z light curves. The field of iPTF13z is shown in Fig. \ref{fig:finder13z}.

We checked our P48 $R$-band photometry against the PTFIDE (PTF Image Differencing \& Extraction) pipeline \citep{masci17} photometry, which was computed in the 0--520-day interval. The mean difference between the PTFIDE output and the output from FPipe was $\Delta~m_{PTFIDE-FPIPE}~=~-0.057\pm0.069$ mag. As PTFIDE is only available for the P48 data, but FPipe is available for both data from P48 and P60, we used the latter for consistency.

The early $R$-band photometry from P48 is presented in Table~\ref{p48r13z_early}. Later photometry from P48 and P60 is presented in Table~\ref{13zp60P48}.

PTF coverage of the field iPTF13z field began with the P48 on 2009 March 17 (3.9 years before discovery) and was continued until 2012 July 28. This multi-epoch coverage allowed us to search for pre-explosion outbursts of the progenitor star, which have proven fruitful in other SN~IIn cases \citep{ofek13a, ofek14outb}. The PTF pre-discovery photometry is analysed in Sect. \ref{sec:outburst}.

Two space-based telescopes also observed iPTF13z. The UltraViolet and Optical Telescope (UVOT) aboard Swift (trigger 32921) observed this SN on 2013 August 28 (207 days). The Spitzer Space Telescope (program 11053, P.I. Fox) observed it on 2015 May 13 (831 days). In neither case was the SN detected.

\subsection{Spectroscopy\label{sec:spec}}
We obtained optical spectroscopy of iPTF13z using the telescopes and spectrographs listed in Table \ref{speclog}, where details of the observations also are given. The spectra were reduced with standard pipelines for each combination of telescope and spectrograph. A spectral sequence is shown in Fig. \ref{fig:spec13z}. The spectra are available via the WISeREP\footnote{Weizmann Interactive Supernova data REPository (\url{http://wiserep.weizmann.ac.il})} database \citep{yaron12}.

\section{Analysis\label{sec:analysis}}

\subsection{Distance and extinction\label{sec:distext}}
Assuming a $\Lambda$CDM cosmology with $H_0 = 70.0$ $\rm km$ $\rm s^{-1}$ $\rm Mpc^{-1}$, $\Omega_M = 0.279$, and $\Omega_{\Lambda} = 0.721$ from the nine-year WMAP observations \citep{hinshaw13} and redshift $z~=~0.0328$ from host galaxy emission lines (Sect. \ref{sec:specanalysis}) gives the luminosity distance $d_L = 144.1$ Mpc and distance modulus $\mu = 35.79$ mag for iPTF13z.

The Milky Way (MW) colour excess towards iPTF13z is $E(B-V)_{MW} = 0.063$ mag \citep{schlafly11}\footnote{Accessed via the NASA/IPAC Extragalactic Database, NED.}. This gives $A_V = 0.20$ mag assuming $R_V = 3.1$ \citep{fitzpatrick99}.

The \ion{Na}{I} D absorption doublet is commonly used to estimate the extinction in SN host galaxies using SN spectra. From the classification spectrum of iPTF13z, obtained at 17~days (see Sect. \ref{sec:spec}) we measure a limit of EW(Na~I~D) $\leq 0.82$ \AA\, based on the most prominent feature around the \ion{Na}{I}~D wavelength region; however, we do not consider this a clear detection of \ion{Na}{I}~D. Using the relation by \citet*[eq. 9]{poz12}, the limit on the EW implies $E(B-V)_{host} < 0.13$ mag. Examining the MW absorption from \ion{Na}{I} D at $\lambda_{obs} = 5893$ \AA\ we find EW(Na~I~D)$_{MW} = 0.81$ \AA\, and know that $E(B-V)_{MW} = 0.063$ mag. Assuming the dust properties in the MW and the SN host to be similar, this suggests that $E(B-V)_{host} \lesssim 0.063$ mag. We can also check our host extinction estimate using the Balmer decrement. For conditions typical in \ion{H}{II} regions, it is $\frac{F_{H\alpha}}{F_{H\beta}} = 2.86$ \citep{osterbrock89}. The SN spectrum obtained at 958 days (Sect. \ref{sec:specanalysis}) appears dominated by host galaxy light. After correcting for MW extinction, from Gaussian profiles fitted to the Balmer lines we found the flux ratio $\frac{F_{H\alpha}}{F_{H\beta}} = 2.5$. A Balmer decrement $< 2.86$ indicates no host extinction. In summary, the results above suggest that the extinction from the host of iPTF13z is negligible. Therefore, in this paper we only correct for MW extinction.

\begin{figure*}
   \centering$
    \includegraphics[width=16.9cm,angle=0]{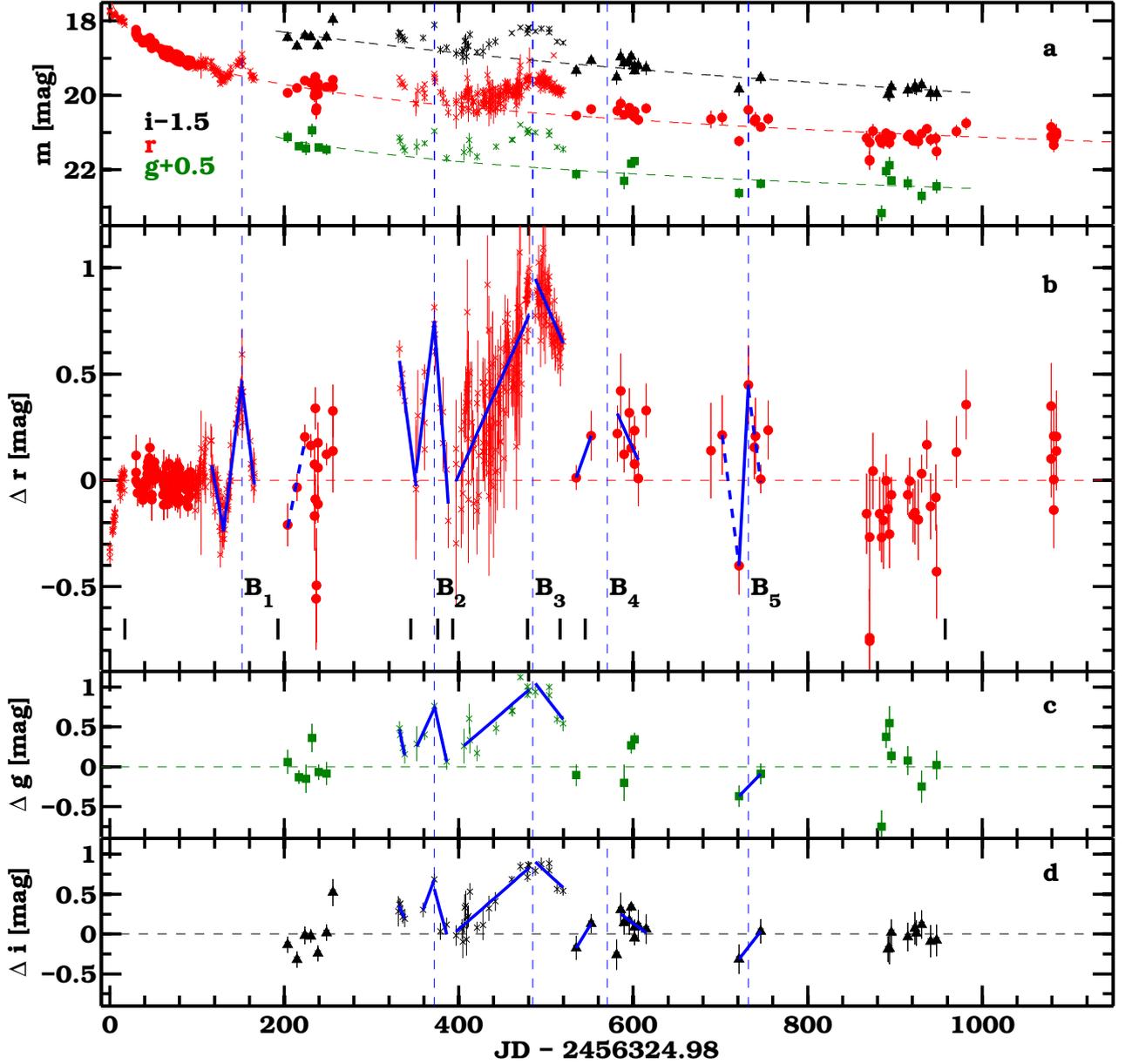}$
    \caption{Light curves of iPTF13z with power-law fits to the $gri$ band declines (panel a) and after removal of the power-law fits (panels b, c, and d). Filled symbols are used for the power-law fits, $\times$ symbols were excluded from the fits. The dashed vertical blue lines indicate the epochs of bump maxima. The  bumps, B$_{1-5}$, are labelled panel b. The black tick marks in panel b show the epochs of the spectra. Bumps and undulations present in all the filters have been marked with unbroken bold blue lines. Bumps and undulations seen only in the $r$ band are marked with dashed  bold blue lines.}
    \label{fig:13zanalysis}
\end{figure*}

\subsection{Photometry}

\subsubsection{Light curves and their bumps\label{sec:LC}}
The $Bgri$ light curves of iPTF13z are shown in Fig. \ref{fig:lc13z}. The brightest observed magnitude of iPTF13z was $M_{r}~=~-$18.2 mag. The SN was followed only in the $R$ band with P48 until $\sim$200~days, then $g$, $r$, and $i$ band follow-up started with P60. A few points in the $B$ band were also obtained at $\sim$600~days and between 700 and 800~days. We took into account the distance modulus and the $r$-band extinction discussed in Sect.~\ref{sec:distext} to compute the absolute $r$-band magnitude scale shown by the right-hand y-axis in Fig.~\ref{fig:lc13z}.

For the purpose of matching the P48 Mould $R$-band and P60 $r$-band measurements, the P48 light curve was shifted by $S = 0.26$ mag. The value of $S$ was computed by averaging the difference between the P48 light curve interpolated over the P60 $r$-band epochs and the corresponding P60 $r$-band photometry. This shift accounts for the different templates used for the host subtraction of the P60 and P48 images, and for the different filter profiles in the two telescopes. This shift has been used in all the following analyses and figures, but the original magnitudes are reported in Tables~\ref{p48r13z_early} and \ref{13zp60P48}.

The light curves of iPTF13z were monitored starting after peak brightness; therefore, they evolve to fainter values from discovery onwards. After the first 35~days at least, the decreasing trend of the $r$-band light curve can be well represented by a power law (PL). This is shown in Fig.~\ref{fig:13zanalysis}a, where the power-law fit is shown by a red dashed line. The $r$-band light curve declined rapidly at early epochs and more slowly at later epochs. A decline of 1.7 mag was observed in the first 150 days, but the light curve took until 1000 days to fade by the same amount. A PL is a good approximation of the decline of the $g$ and $i$ bands as well, and we show the best fits to these light curves with green and black dashed lines in Fig.~\ref{fig:13zanalysis}a.

The  PL fits were made only to a portion of the light curves (see filled symbols in Fig.~\ref{fig:13zanalysis}), excluding those parts where  clear deviations from the decline were observed. We see that undulations and episodes of re-brightening strongly characterise the light curves of iPTF13z. When we subtract the PL fit from the light curves of iPTF13z, as shown in Figs.~\ref{fig:13zanalysis}b, c, and d, these bumps clearly emerge \citep[see][for a similar analysis]{martin15}. The PLs fitted here (e.g. $r(t) \propto t^{-0.045}$ for the $r$ band) to light curves in the time-magnitude domain are used for decline removal to make the bumps easier to study. In Figs. \ref{fig:13zanalysis}b, c, and d, $\Delta m$ for a certain epoch is computed as the magnitude from the fit to the decline minus the magnitude from the photometry.

The $r$ band has the best coverage and sampling, thanks to the P48 observations. In Fig.~\ref{fig:13zanalysis}b we identify and name the bumps occurring in this band and check if they are also detected in the other bands. The properties of the bumps are given in Table \ref{13z:bumps}. The first conspicuous bump, B$_1$, occurred between $115 - 165$~days. Here the PL-subtracted light curves first linearly declined by 0.22 mag in 14 days, then rose to 0.48 mag above the trend of the decline over the next 21 days, and returned to the decline trend level over the next 14 days. The three phases of bump B$_1$ are shown by solid blue lines in Fig.~\ref{fig:13zanalysis}b, which are best fits to the data at the above-mentioned phases. The maximum of B$_1$, occurring at 152 days, is marked by a vertical dashed blue line. The dense light curve sampling and the small photometric errors at this phase make the identification of this bump quite certain, even though we cannot confirm it also happened in the other bands, due to lack of coverage.

After 165~days, the first well-sampled portion of light curve ended. In the period between  200 and 260 days, some scattered epochs of photometry are available. This photometry was taken at higher air mass than the photometry prior to 165 days, due to the approaching conjunction with the Sun. There was a clear rise from $\Delta r = -$0.21 mag to $\Delta r =$ 0.20 mag between 200 and 225 days, but this was not confirmed in the other bands, so we marked it with a dashed blue line.

Between 330 and 390  days, the $r-$band photometry shows a new bump or undulation, again highlighted with linear fits in Fig.~\ref{fig:13zanalysis}b. In this phase, for 20 days the decline trend subtracted light curves dropped by 0.57 mag to the decline trend level, then it rose by 0.75 mag over the following 20 days, reaching peak at 372 days. Finally, over the next 14 days, it dropped to $\Delta r = -$0.1 mag. This bump, B$_2$, was clearly also observed  in the $g$ and $i$ bands, as highlighted in  Figs.~\ref{fig:13zanalysis}b and c, with similar time scale and magnitude variation.

We note that the time scale and amplitude of bump B$_2$ was  similar to those observed for bump B$_1$. On the contrary, between 400 and 520  days, a broader and higher bump, B$_3$, occurred in the $r$ band. It slowly rose from $\Delta r = 0$ mag up to $\Delta r = 0.9$ mag (at 485 days, see blue dashed vertical line), then it declined faster over the following 35 days before light curve coverage was interrupted at 520 days. Bump B$_3$ was also clearly observed in the $g$ and $i$ bands.

Later than 520 days, we have four  scattered clusters of photometry (three in $g$ and one in $i$). At this phase the photometry is poorer in coverage and uncertain; however, we can identify two bumps in the $r$ band peaking at 570 days (B$_4$) and 732 days (B$_5$), at 0.40 and 0.44 mag above the decline trend. Bump B$_4$ was confirmed in the $i$ band (not in the $g$ band, due to poor coverage), whereas the rise of B$_5$ was observed in both the $g$ and the $i$ band. Their durations were $\approx$70 and $\approx$45 days, respectively. Bump B$_5$ showed a shape similar to the shape of B$_1$ and B$_2$.

\begin{figure}[h!]
   \centering
    \includegraphics[width=9cm]{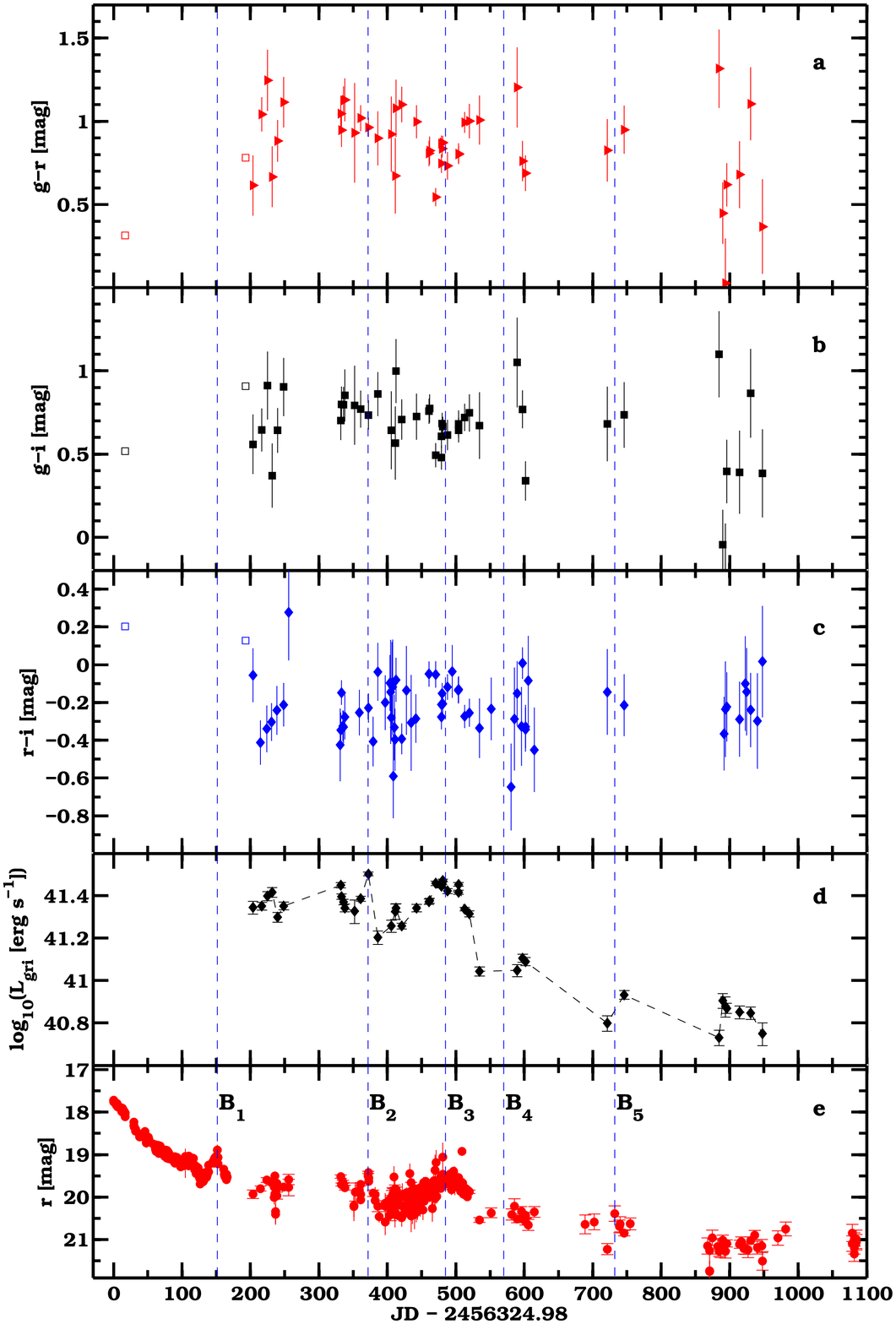}
    \caption{Panels a-c: Colours of iPTF13z from the P60 telescope, corrected for galactic reddening, shown with filled symbols. Colours obtained from synthetic photometry on spectra from 17 and 193 days are shown as open square symbols. Panel d shows the pseudo-bolometric light curve ($gri$) for iPTF13z, with the $r$-band light curve shown for comparison in panel e (labelled bumps B$_{1-5}$). Epochs of the bumps are marked with vertical dashed blue lines.}
    \label{fig:colbol}
\end{figure}

\subsubsection{Colour evolution and pseudo-bolometric light curve \label{sec:colour}}
Photometric colour information is only available for iPTF13z later than 200 days (Fig. \ref{fig:colbol}). As shown in Figs.~\ref{fig:lc13z} and ~\ref{fig:13zanalysis}, the $g$, $r$, and $i$ light curves evolved in a similar way, especially in terms of bumps. This is reflected by the absence of a strong colour evolution, as shown in Fig. \ref{fig:colbol}a-c for three different colours. The $g-r$, $g-i$, and $r-i$ colour evolution appear rather flat across the covered time range, with average values $<g-r> ~=~ 0.87\pm0.24$ mag, $<g-i> ~=~ 0.66\pm0.24$ mag, $<r-i> ~=~ -0.23\pm0.15$ mag. This can be compared to the late colour evolution of the long-lasting SN~IIn 1988Z \citep{aretxaga99, pastorello02} in $(V-R)$ and $(B-V)$, which showed no major trend between 400 and 1200  days after discovery. For completeness, we obtained colours by synthetic photometry (using \texttt{synphot} by \citealp{ofek14c}) on the spectra from 17 and 193 days, de-reddened and flux calibrated. These colours are plotted in Fig. \ref{fig:colbol}, but are not included in the average values given above. Evolution from bluer to redder synthetic colour can be seen in the $(g-r)$ and $(g-i)$ synthetic colours of iPTF13z.

Using the $g$, $r$, and $i$ photometry from P60, we constructed a pseudo-bolometric light curve at the phases when colour information is available ($> 200$ days). The pseudo-bolometric luminosities were estimated by interpolating the $r$ and $i$ bands to the epochs of the $g$ band, converting the obtained magnitudes into specific fluxes at their respective effective wavelengths, and integrating the spectral energy distributions (SED) in the wavelength range between the $g$ and $i$ band effective wavelength. The errors were estimated by simulating 1000 artificial SEDs for each epoch, based on the fluxes and the flux errors. The pseudo-bolometric luminosity is presented in Fig. \ref{fig:colbol}d and compared to the $r-$band light curve (Fig. \ref{fig:colbol}e). We can clearly identify bumps B$_2$, B$_3$, and B$_5$ in the pseudo-bolometric light curve.

We can approximate the electromagnetic energy output of iPTF13z using a luminosity estimate based on our $R/r$-band photometry. For this purpose we use the expression
\begin{equation}
\label{eq:lum}
L(M_r) = 3.04\cdot10^{35} \times 10^{(-0.4 M_r)} \rm  ~erg ~s^{-1}
\end{equation}
(based on solar bolometric values) to convert $r$-band magnitudes into bolometric luminosity values. The irregular spectral coverage of iPTF13z (especially during the early bright phase) makes it difficult to determine the bolometric correction as a function of time, so for our estimate we assume the constant bolometric correction to be zero \citep[see][]{ofek14outb}. Integrating under a piecewise fit of first- to third-order polynomials to the $0-1100$ days $r$-band light curve using Eq. \ref{eq:lum} for $L$ gives the energy estimate $6 \times 10^{49}$ erg. The behaviour of iPTF13z during its rise to peak brightness, and around peak, is unknown. This makes it difficult to determine the total bolometric output of the SN, but the above estimate gives the lower limit $\gtrsim 6\times10^{49}$ erg.

\begin{figure}[h!]
   \centering
    \includegraphics[width=9cm]{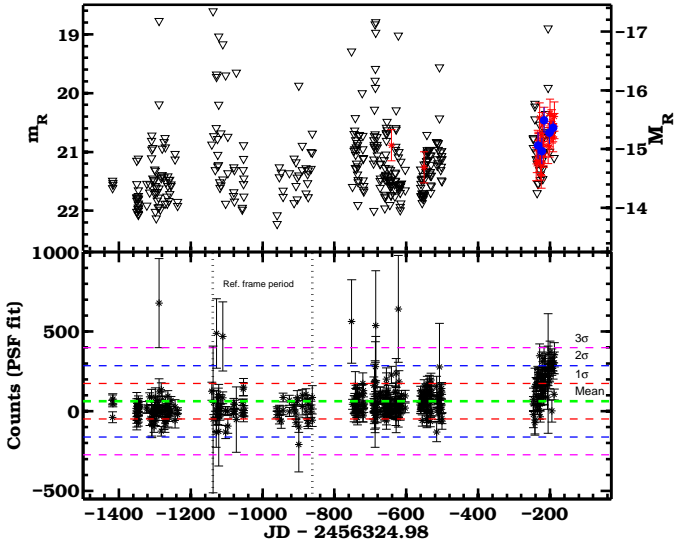}
    \caption{Pre-discovery light curve for the position of iPTF13z, from P48 (Mould $R$-band). The upper panel shows the light curve;  triangles indicate upper limits from the PSF photometry and red symbols detections. The filled blue circles are binned representations of the detections (10-day bins). The lower panel shows the same measurements in counts, with dashed horizontal lines showing the mean (green) and standard deviations ($1\sigma$, red; $2\sigma$, blue; $3\sigma$, purple) of the counts from all pre-discovery observations. The dashed vertical lines show the period from which the frames for the reference stack were taken.}
    \label{fig:prediscLC}
\end{figure}

\begin{figure*}
   \centering$
    \includegraphics[width=18cm,angle=0]{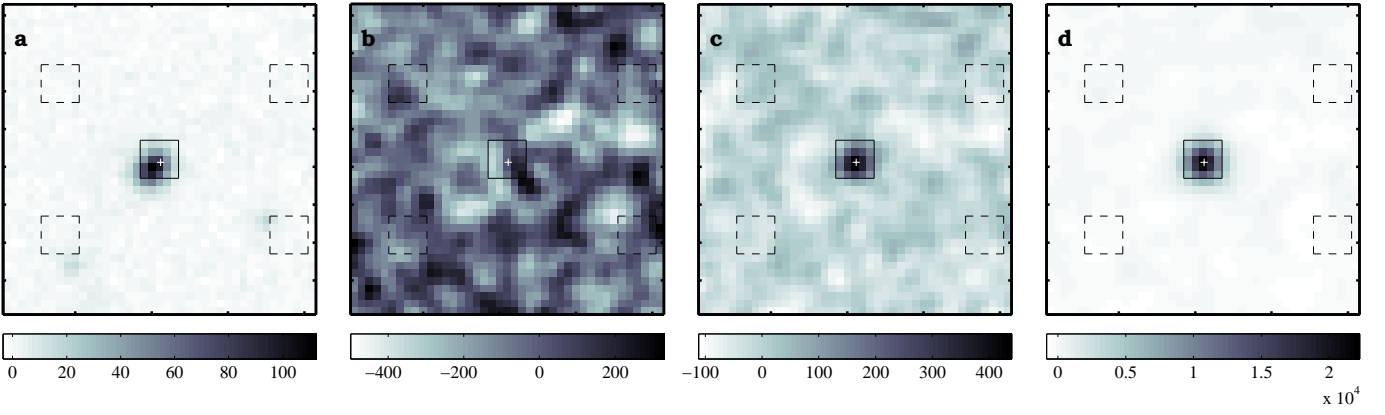}$
    \caption{Stacks of Mould $R$-band images from the P48 telescope. \textbf{a:}  Stacked reference image of the iPTF13z host galaxy. \textbf{b:} The 332 earliest frames (from $-1417$ to $-242$ days) stacked, giving signal-to-noise ratio $S/N = 1$ at the site of the SN. \textbf{c:} Stack of the 54 frames from from $-238$ to $-188$ days, with $S/N = 15$, showing the pre-discovery outburst. \textbf{d:} All  456 post-discovery frames (from 0 to 870 days) stacked, with $S/N = 60$, showing the SN. The scale of the axes is in pixels ($1.01\arcsec / \rm pixel$). The field of view is $41\times41$ pixels, with north down and east to the right. The aperture for the SN location is shown as a box with whole sides. Apertures for the background are shown as dashed boxes. These boxes were used to estimate the signal and the noise in the stacks, to calculate the S/N. The white cross shows the PSF based position of the outburst candidate. Intensity scales, in units of counts, are given below each subplot.}
    \label{fig:outburst}
\end{figure*}

\subsubsection{A pre-discovery outburst \label{sec:outburst}}
Outbursts before the explosions have been detected for a few SNe~IIn, for example in the cases of SNe~2009ip \citep[e.g.][]{pastorello13, mauerhan2013a}, 2010mc \citep{ofek13a}, 2011ht \citep{fraser13_11ht}, LSQ13zm \citep{tartaglia16}, 2015bh \citep{thone17, ofek16}, and in the sample presented by \citet{ofek14outb}. It is therefore of interest to look for the presence of an outbursting progenitor also in the case of iPTF13z. We searched for pre-discovery outburst events for iPTF13z using our 386 archival P48 $R$-band images of the field from 2009 March 17 to 2012 July 28.

For the pre-explosion images, host subtraction was done in a way similar to that described in Sect. \ref{sec:phot}. The same stacked reference frame as in Sect. \ref{sec:phot} was used. Figure \ref{fig:prediscLC} (top panel) shows that the deepest limits are in that time range, and that the counts in that interval are constant and on average lower than in other portions of the pre-discovery light curve. Stacking by median combining and sigma clipping, using the \texttt{sim\_coadd} routine by \citet{ofek14c}, of the selected images was done to build the reference image. 

The stacking of the images was done for a $41\times41$ pixel subframe around the SN site from each host and sky subtracted image. The reference image is shown in Fig.~\ref{fig:outburst}a. We note that the brightest pixel in the host galaxy image is at a different location than the SN position (marked by a white cross). In order to look for outbursts before the explosion of iPTF13z, the pre-discovery, host subtracted images were first divided into two time intervals: one with 332 frames (from $-1417$ to $-242$ days) and one with 54 frames (from $-238$ to $-188$ days). The latter set of frames was selected since the pre-discovery light curve suggested that a signal was present at the SN location during this 50-day period. This is clearly seen in the bottom panel of Fig.~\ref{fig:prediscLC}, where several points around $-210$ days are between 2$\sigma$ and 3$\sigma$ above the average pre-discovery counts.

The host subtracted images in these two ranges were stacked, and from them the signal from the SN site was measured using a $5\times5$ pixel aperture centred on the SN site (at 1 pixel resolution) as determined via iPTF astrometry. Four $5\times5$ pixel apertures in the vicinity of the SN site were selected to quantify the noise. The two stacked images corresponding to the two different intervals are shown in Figs.~\ref{fig:outburst}b and \ref{fig:outburst}c. In Fig. \ref{fig:outburst}d, we show a stacked image of all the post-discovery frames, with the expected clear detection of the SN. 

When compared to Figs.~\ref{fig:outburst}b and \ref{fig:outburst}c, it is clear that no outburst is detected in the first time interval (in the SN region, at the centre of Fig.~\ref{fig:outburst}b, the signal is comparable to the noise), while in the time interval around $-210$ days there is a clear signal. As we applied the World Coordinate System (WCS) from the reference image to all the subtracted frames, this allowed us to compare the location of the outburst candidate (Fig.~\ref{fig:outburst}c) to that of the actual SN (Fig.~\ref{fig:outburst}d). A 2D Gaussian function was fit to determine the PSF centroids of the two sources. The distance between the SN and the outburst candidate was found to be negligible ($0.064\pm0.065$ pixels, $1\sigma$ error), confirming the coincidence in location.

Considering the $R$-band light curve obtained before discovery at the iPTF13z site (Fig. \ref{fig:prediscLC}, top panel) and the strong detection in the stacked images around $-$210 days, it appears that the progenitor star of iPTF13z had a $\approx 50$-day outburst (red stars in the upper panel of Fig. \ref{fig:prediscLC}) around 210 days before the discovery of the SN. The limiting magnitude of the unfiltered CRTS images from $-137$ days and $-129$ days (Sect. \ref{sec:disc}) corresponds to $M \lesssim -17$ mag, suggesting that the SN had not exploded at that time, but providing no further information about any precursor activity. We cannot exclude that the pre-discovery outburst is the start of the SN rise, but a $\approx 200$-day rise time for a Type IIn SN would be almost unprecedented; SN 2008iy rose for $\approx 400$ days \citep{miller10}. 

From the 10-day binned average magnitudes in the light curve (blue circles), we find that the average absolute magnitude of this outburst is $M_R \approx -15$. Assuming zero bolometric correction and using Eq. \ref{eq:lum} this corresponds to a luminosity of $L_{prec} \approx 3\times10^{41} \rm ~erg ~s^{-1}$. For the 50-day duration, this gives an energy output of $E_{prec} \approx 1\times10^{48} \rm ~erg$, if we assume that the outburst had  constant luminosity during the covered period.

\subsubsection{Photometric comparison to other SNe~IIn\label{sec:complight}}

\begin{figure*}
   \centering
    \includegraphics[width=15cm,angle=0]{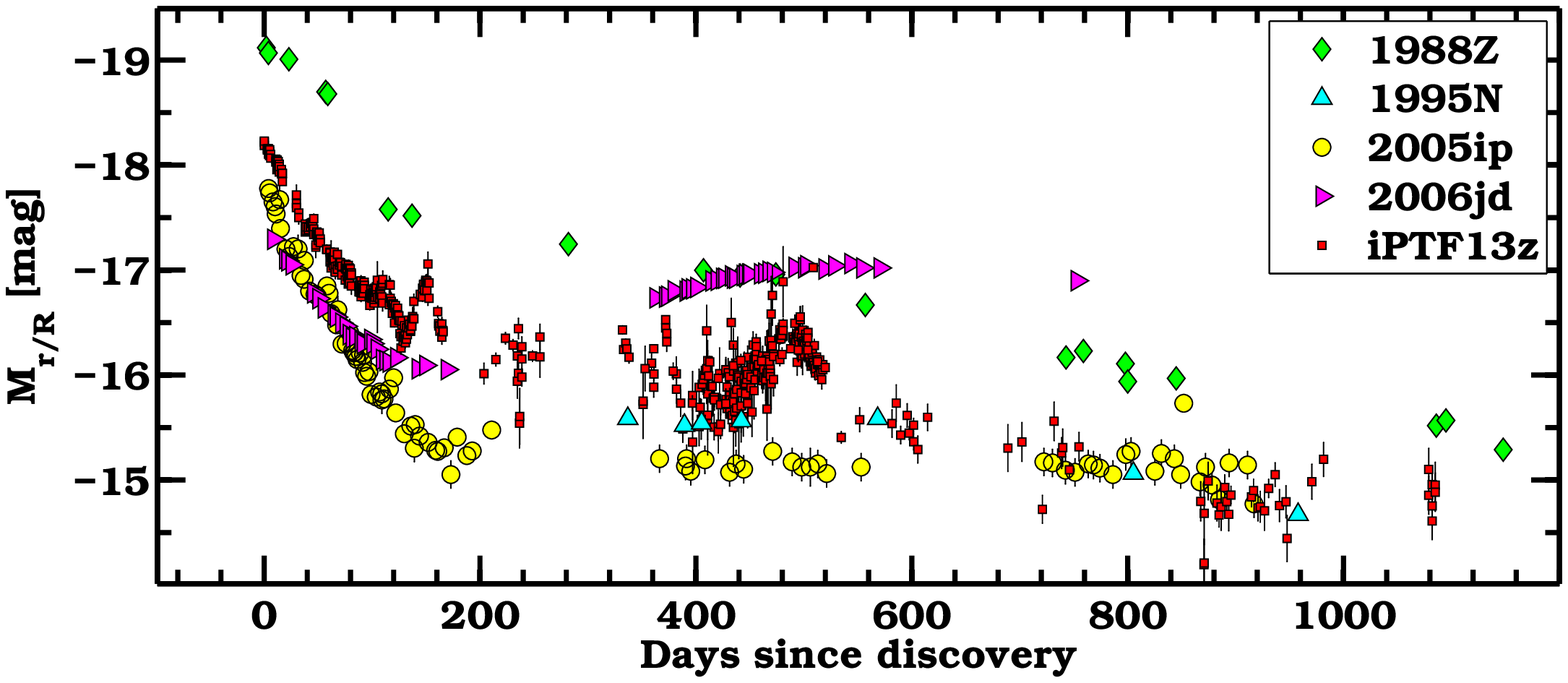}\\    
    \caption{\label{fig:complc}Light curve of iPTF13z compared to those of other long-lasting SNe IIn. Photometry of SN 1988Z from \citet{aretxaga99}, SN~1995N (from \citealp{pastorello05}, adopting the SN age in that paper), SN~2005ip from \citet{stritzinger12} and \citet{smith09_2005ip} and SN~2006jd from \citet{stritzinger12}.}
\end{figure*}

We will now compare the light curve of iPTF13z to other relevant SN~IIn light curves from the literature. In Fig. \ref{fig:complc}, we plot SN~IIn light curves in the $r/R$ band from the literature, choosing events resembling iPTF13z. These transients have been followed for more than 1000 days. All of them were discovered after peak brightness. The behaviour of iPTF13z resembled that of  several long-lasting SNe~IIn, whose prototypical event was SN 1988Z \citep{turatto93}. In our comparison, we also include SN~1995N \citep{fransson02, pastorello05}, SN~2005ip \citep{smith09_2005ip, fox09, stritzinger12}, and SN~2006jd \citep{stritzinger12}. All these SNe IIn showed a steep initial decline until $\sim 100-200$ days, which was followed by a long slower decline. The maximum observed magnitude of iPTF13z in the $r$ band was slightly brighter ($\lesssim -$18.3 mag) than those of SNe~2005ip and 2006jd, which did not reach $-$18 mag. SN~1988Z was more luminous, peaking at $-19$ mag, whereas SN1995N was only observed during its slow declining phase. The typical $M_{r/R}$ during the late phase ranges from $\sim-$17 (SN 1988Z) to $\sim-$15 (SN 2005ip) mag.

Undulations and bumps in SN~IIn light curves are rare but have been observed in a few cases. Among these SNe, both iPTF13z and SN~2006jd showed a pronounced bump peaking at $\sim$500 days. However, the bump in SN~2006jd was only visible in the $r$-band light curve, whereas that of iPTF13z was present in all the observed bands. The bump in SN~2006jd was longer than that of iPTF13z and lasted from 300 days to 700 days, with a maximum amplitude of $\sim 1$ mag (similar to that of iPTF13z) at 544 days. For SN 2006jd, the increase in $r$ brightness coincided with an increase in H$\alpha$ luminosity \citep{stritzinger12}.

During the decline of SN 2009ip, after its major brightening (called the 2012B event), the fading was interrupted by a re-brightening in visible light lasting $\sim 10$ days and reaching an amplitude of $0.2$ to $0.5$ mag in the optical \citep{fraser13a}. The data presented by \citet{graham14} showed that the bump started $\approx 26$ days after peak in all the optical bands. Bump B$_1$ in the light curve of iPTF13z, which was monitored only in the $r$ band, had a larger amplitude and occurred on a longer time scale than that of SN 2009ip. The same is true for the other bumps observed in iPTF13z (Fig.~\ref{fig:13zanalysis}). The main bump of SN 2009ip occurred  in all the bands and was characterised by colours slightly evolving to bluer values (see $g-r$ in \citealp{graham14}, their Fig.~9). Like bump B$_3$ of iPTF13z, the main bump of SN 2009ip occurred in all observed photometeric bands. \citet{martin15} fitted and subtracted the main trend from the declining light curve of SN 2009ip to bring out the fluctuations and noted that the optical light curves showed other low-amplitude and short bumps at $-20$, $+15$, and $+30$ days from the main bump. For SN 2009ip, \citet{martin15} concluded that the fluctuations in the decline were likely driven by changes in the continuum rather than variations in the emission lines (see Sect. \ref{sec:Halpha}).

SN 2010mc \cite[= PTF10tel, ][]{ofek13a} displayed the beginning of a bump as a noticeable rise relative to the decline of its $R$-band light curve at 40 days after maximum. The light curve was then broken, making it hard to constrain the properties of this bump. The SN~IIn/Ibn hybrid SN 2005la \citep{pastorello08} showed a bump lasting merely 4 days in its $R$-band light curve, around 18 days after $R$-band maximum \citep{pastorello08}.

\begin{figure*}
   \centering
        \includegraphics[width=15cm,angle=0]{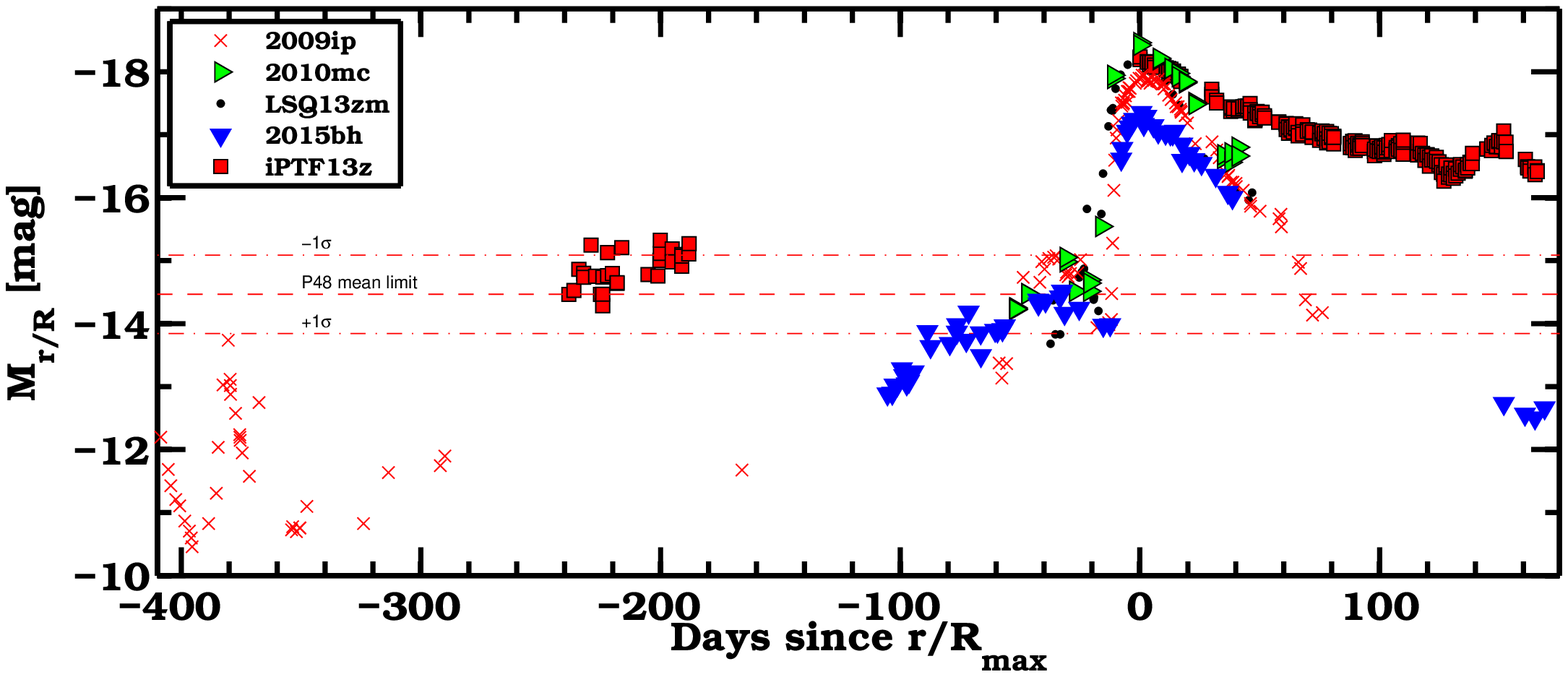}
  \caption{\label{fig:outburst_comp} Comparison of pre-explosion outbursts for SNe IIn. The $r/R$-band light curves of iPTF13z compared to SN 2010mc \citep{ofek13a}, LSQ13zm \citep{tartaglia16}, and SN 2015bh \citep{eliasrosa16}, along with the 2012B outburst of SN 2009ip \citep{pastorello13, fraser13a, fraser15}. For the plotting, the peak of iPTF13z has been assumed to occur at 0 days. For comparison, the red dashed line shows the mean limit of P48 detections in absolute magnitudes for the distance and MW extinction of iPTF13z.}
\end{figure*}

Another important characteristic of iPTF13z is the detection of a rather luminous outburst before the main explosion (at $-$210 days). Other SNe IIn also showed this behaviour. Among the precursors studied by \citet{ofek14outb}, we note a similarity in absolute magnitude between the precursors of \object{PTF10bjb} and iPTF13z. SN 2009ip was observed outbursting in 2009, in 2011, and finally in 2012 before its main event, 2012B \citep{pastorello13}. The pre-max light curves of SN 2009ip and that of iPTF13z are shown in Fig.~\ref{fig:outburst_comp}. The precursor of iPTF13z (around $-15$ absolute magnitude) is brighter by $\sim$2 mag than the average activity of SN 2009ip before 2012. Other SNe such as SN 2010mc \citep{ofek13a, ofek14outb}, LSQ13zm \citep{tartaglia16}, and SN 2015bh \citep{thone17, ofek16, eliasrosa16} exhibited pre-explosion outbursts around $-$14 absolute magnitude, similar to the 2012A event of SN 2009ip. Their light curves are also shown in Fig.~\ref{fig:outburst_comp}. In the cases of SN 2009ip, LSQ13zm, and SN 2015bh, it is not certain that core-collapse of the progenitor star occurred.

\subsection{Spectroscopy\label{sec:specanalysis}}
\begin{figure*}
   \centering$
    \includegraphics[width=16.9cm,angle=0]{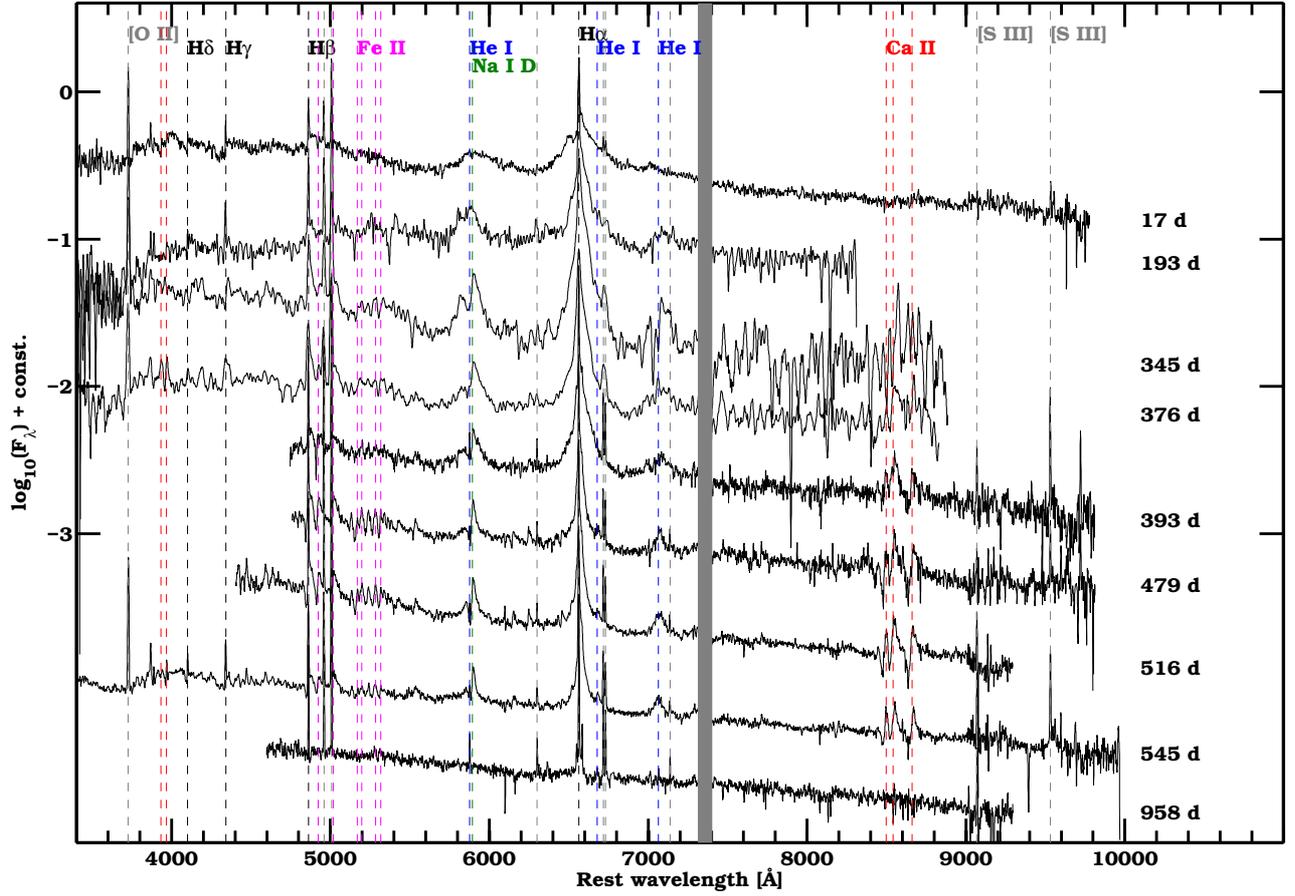}$
    \caption{Spectra of supernova iPTF13z shown at rest wavelength, with line identifications. The spectra have been smoothed to aid the clarity of the plot. Atmospheric spectral features are masked with a grey stripe. The epochs of the spectra are given to the right. Line identifications are given at the top (grey indicates host galaxy lines, other colours SN lines).}
    \label{fig:spec13z}
\end{figure*}

Our nine best spectra of iPTF13z are shown in Figure~\ref{fig:spec13z}. The noisy spectrum from 450 days and the host dominated spectra from 723 days and 958 days have been excluded from the plot. The spectra have been calibrated to the $r$-band photometry and corrected for extinction (Sect.~\ref{sec:distext}).

By fitting host galaxy emission lines ([\ion{O}{iii}]~$\lambda\lambda$4959, 5007; [\ion{S}{ii}]~$\lambda\lambda$6717, 6731; and [\ion{S}{iii}]~$\lambda$9069) with Gaussians, we obtained their observed centroids and computed the redshift ($z~=~0.0328\pm0.0001$) used for iPTF13z in this work.

The optical spectroscopy coverage of iPTF13z was uneven. During the first 200 days after discovery, two spectra were obtained: at 17 days (classification spectrum) and at 193 days. Unfortunately, no spectra were taken during bump B$_1$. Before, during, and right after bump B$_2$, three spectra were obtained at 345, 376, and 393 days. Two more spectra (at 479 and 516 days) were obtained during bump B$_3$, and one (545 days) before the peak of bump B$_4$. Two host galaxy dominated spectra were obtained, at 723 and 958 days.

The classification spectrum has features conforming with the spectroscopically based definition of a SN~IIn by \citet{schlegel90}, showing a blue continuum and having a narrow H$\alpha$ line with a broad base and no broad   P Cygni absorption (which SNe~IIP/L display). The dominating feature of the SN spectrum at all epochs was H$\alpha$ emission.

This SN~IIn was also characterised by the presence of strong \ion{He}{i} $\lambda\lambda$5876, 7065 emission  since the first spectrum. Starting at 345 days, the 5876 \AA\ line shape was affected by the presence of a relatively narrow \ion{Na}{i}~D P Cygni profile. The combination of the two lines produced a double peaked feature.

From 345 days (and clearly from 393 days) to 545 days, the spectra showed the \ion{Ca}{II} near-IR triplet at $\lambda\lambda$8498, 8542, and 8662. The triplet exhibited narrow P Cygni profiles.

In the blue part of the spectrum, a number of \ion{Fe}{II} lines from 4924 \AA\ to 5316 \AA\ showed relatively narrow P Cygni profiles (FWHM of a few $10^2 \rm ~km ~s^{-1}$), clearly visible from 345 days. P Cygni profiles of similar broadness were exhibited by H$\beta$ (and when discernible, H$\gamma$ and H$\delta$) and these were visible since the first spectrum.

The galaxy narrow emission lines contaminated the SN spectrum. The four spectra with sufficient wavelength coverage all showed the host galaxy [\ion{O}{II}] $\lambda$3727 line. Narrow Balmer lines were visible in all the spectra, partly due to SN emission and partly to the galaxy (see Sect.~\ref{sec:Halpha}). The [\ion{O}{iii}] galaxy lines at $\lambda\lambda$4959, 5007 were very bright, as were the [\ion{S}{ii}] $\lambda\lambda$6717, 6731 host galaxy lines. At 958 days, the H$\alpha$ from the SN had become sufficiently weak to reveal the narrow [\ion{N}{ii}] $\lambda\lambda$6548, 6583 lines from the host galaxy, which thus contaminated the H$\alpha$ emission in the previous spectra. Narrow \ion{He}{i}~$\lambda$5876 and [\ion{O}{i}]~$\lambda$6300 were also visible.
 
The continuum of the spectra appeared blue at all epochs (except at 193 days) and did not evolve much. This is in agreement with the relatively constant colours of the SN (Fig.~\ref{fig:colbol}), and  with the colour of the star-forming host galaxy (Sect. \ref{sec:hostgal}) that contaminates the spectra, especially at late epochs.

Overall, the spectra of iPTF13z, with the exception of the first two, were all rather similar. In the following we examine the most important spectral lines in detail, with particular focus on the possible connection between the light curve bumps and the evolution of the spectral features, and on their velocities and fluxes. All the flux measurements are made on photometrically calibrated, extinction corrected, and de-redshifted spectra. The photometric calibration of the spectra is such that the synthetic photometry follows the iPTF13z light curve.

\subsubsection{ H$\alpha$ emission line\label{sec:Halpha}}
Two superimposed Lorentzian profiles were fitted to the continuum-subtracted H$\alpha$ emission line of iPTF13z in order to quantify fluxes and velocities. We first selected two wavelength intervals blueward and redward of the H$\alpha$ emission, where we fit the spectrum with a first-order polynomial in order to reproduce the continuum, which was then subtracted from the spectrum.

\begin{figure*}
   \centering$
    \includegraphics[width=16.9cm,angle=0]{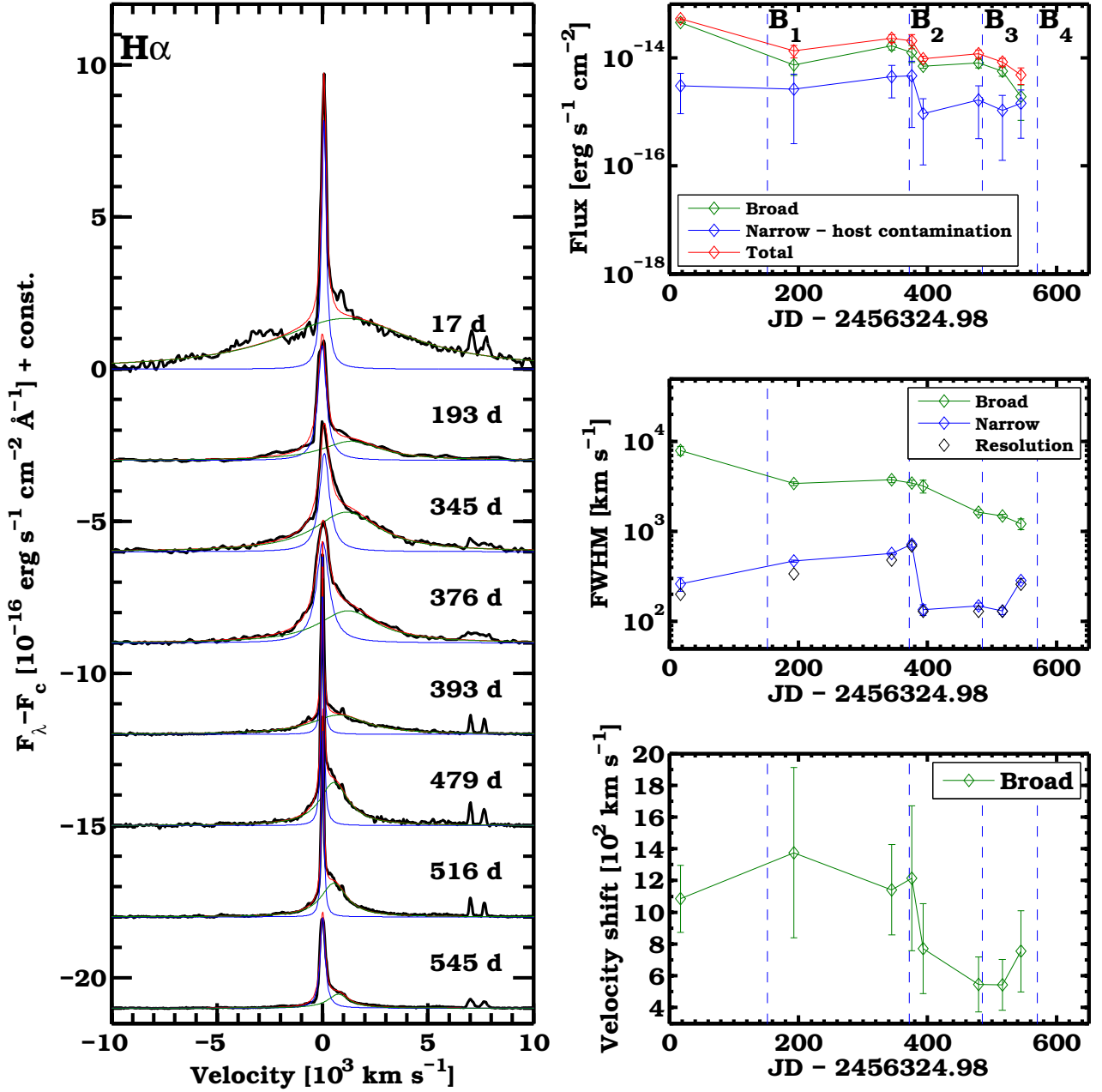}$
    \caption{Fluxes, FWHM velocities, and velocity offsets between the broad and narrow component for the H$\alpha$ line of iPTF13z. The values were obtained by least-squares fits of a two-component Lorentzian profile to spectra calibrated to photometry, de-reddened using $E(B-V)_{MW} = 0.063$ mag, and continuum subtracted. In the left panel, the broad Lorentzian is shown in green, the narrow component in blue, and their sum in red. The epochs of the spectra are given in days. The emission seen at 7000 $\rm km ~s^{-1}$ and 7700 $\rm km ~s^{-1}$ comes from host galaxy \ion{[S}{II]}. We used the ratio between the flux of \ion{[S}{II]} and the narrow H$\alpha$ in the last spectrum (958 days)  to remove the host galaxy contamination in the narrow emission of H$\alpha$ of all the other spectra. We also show the resolution of each spectrum in the centre-right panel. The vertical dashed blue lines show the epochs of the light curve bump peaks (labelled B$_{1-4}$,  upper right panel). The $1\sigma$ error of the least-squares fits were used as a measure of the statistical errors. The systematic errors were estimated by the difference between the Gaussian and Lorentzian fits to the lines. The error bars in the plots show statistical and systematic errors added in quadrature.}
    \label{fig:13zHalpha}
\end{figure*}

The two Lorentzian profile components were fitted to the emission feature using a least-squares method. We use the $1\sigma$ error of the fit as a measure of the statistical errors of the luminosities and velocities found. Systematic errors were estimated by doing least-squares fitting of Gaussian profiles to the H$\alpha$ line in order to  use the difference between the Gaussian and the Lorentzian fits as a measure of the systematic errors. The errors given in Tables \ref{13z:havel} and \ref{13z:haflux} and in Fig.~\ref{fig:13zHalpha} are the $1\sigma$ errors and systematic errors added in quadrature. One Lorentzian reproduced the broad base of the H$\alpha$ emission, the other Lorentzian fitted the narrow component. The best fits to the continuum-subtracted H$\alpha$ profiles are shown in the left panel of Fig.~\ref{fig:13zHalpha}.

The broad component (green lines and diamonds in Fig.~\ref{fig:13zHalpha}) showed a trend towards lower fluxes with time, following the light curve shape. In correspondence with the rise of bump B$_2$, the flux of the broad H$\alpha$ component rose, then dropped after B$_2$ had vanished (393 days). Again, the flux marginally rose during the rise of bump B$_3$ (479 days), then it dropped in correspondence with the decline of B$_3$ (516 and 545 days). The epochs of the bumps are marked by blue dashed vertical lines in the right-hand panels of Fig.~\ref{fig:13zHalpha}.

The narrow component flux shows a similar trend, with fluxes decreasing with time. When computing the flux of the narrow component of H$\alpha$, we removed the host contamination by the following method: we first measured the ratio between the flux of the \ion{S}{ii} lines and that of the narrow H$\alpha$ emission in the last spectrum (958 days) where the galaxy emission largely dominates; then we measured the \ion{S}{ii} fluxes in the other spectra (again with Lorentzian fits), and computed the host galaxy H$\alpha$ emission in each spectrum by multiplying the \ion{S}{ii} fluxes by the ratio F(H$\alpha$)/F(\ion{S}{ii}) determined from the 958 days spectrum; we then subtracted this H$\alpha$ narrow line flux of the galaxy from the total narrow H$\alpha$ flux to obtain the flux from the SN in that component. The galaxy contamination for this narrow line was typically $\sim 50$ \%.

The H$\alpha$ line luminosities and velocities for the two components are summarised in Tables \ref{13z:havel} and \ref{13z:haflux}. We note that the spectra at 393 days and later were more affected by host contamination in the H$\alpha$ region, due to the presence of [\ion{N}{ii}] lines, and this might partially influence the measurements.

The large fluctuations in the $r$-band light curve of iPTF13z cannot be solely attributed to the observed small variations in the H$\alpha$ flux, but mainly to a change in the continuum flux. As an example, we consider bump B$_3$, which was well covered in both photometry and spectroscopy. The $r$-band amplitude of bump B$_3$ (Table \ref{13z:bumps}) was 0.90 mag, corresponding to a factor 2.3 flux increase in the $r$ band for the rise of the bump. From Table \ref{13z:haflux}, we see that the ratio in H$\alpha$ emission line flux between 393 days  (at the start of bump B$_3$) and 479 days (around B$_3$ peak) is 1.2, i.e. smaller than the flux change in the whole $r$ band. This suggests that changes in the continuum (rather than in line emission) caused the bumps. The bumps also occurred  in the $g$ and $i$ bands where emission lines did not dominate the spectrum. The clear appearance of bumps B$_2$, B$_3$, and B$_5$ in the pseudo-bolometric light curve (Fig. \ref{fig:colbol}d) also indicates that changes in the continuum (rather than in individual emission lines) caused the bumps.

The broad component FWHM showed progressively lower values, from $\sim 10^4$ km~s$^{-1}$ at the early epochs to $\sim 10^3$ km~s$^{-1}$ at $\sim$ 600 days. No sharp variations were seen at the epochs of the bumps. The narrow component was unresolved, with its FWHM tracing the resolution of the spectra (see black diamonds in the centre-right panel of Fig.~\ref{fig:13zHalpha}). We measured the spectral resolution of each spectrum based on the width of the host [\ion{S}{II}] $\lambda\lambda 6717, 6731$ lines (when resolved; otherwise we used the host [\ion{O}{III}] $\lambda 4959$ line, see Table \ref{speclog}).

For all epochs, a redshift of the broad H$\alpha$ component was seen relative to the narrow component, where the latter appeared centred at zero velocity. The velocity shift is plotted in the lower  right panel of Fig. \ref{fig:13zHalpha}. The shift was $\sim 1100-1400 \rm ~km ~s^{-1}$ up to the peak of bump B$_2$. Afterwards, a marked decline to about $600-800 \rm ~km ~s^{-1}$ took place. We note that on the broad red side of H$\alpha$, \ion{He}{i} $\lambda$6678 was present, and might have contaminated the main line, altering the velocity shift (and marginally FWHM) measurements, especially at early epochs.

It is interesting to note that all the properties of the H$\alpha$ emission plotted in the three right panels of Fig. \ref{fig:13zHalpha} show a pronounced change at 400 days. When comparing this spectroscopic development to the multi-band (Fig. \ref{fig:lc13z}) and bolometric (Fig. \ref{fig:colbol}d) light curves, we see that this is the starting epoch of bump B$_3$.

\begin{figure*}
   \centering$
    \includegraphics[width=16.9cm,angle=0]{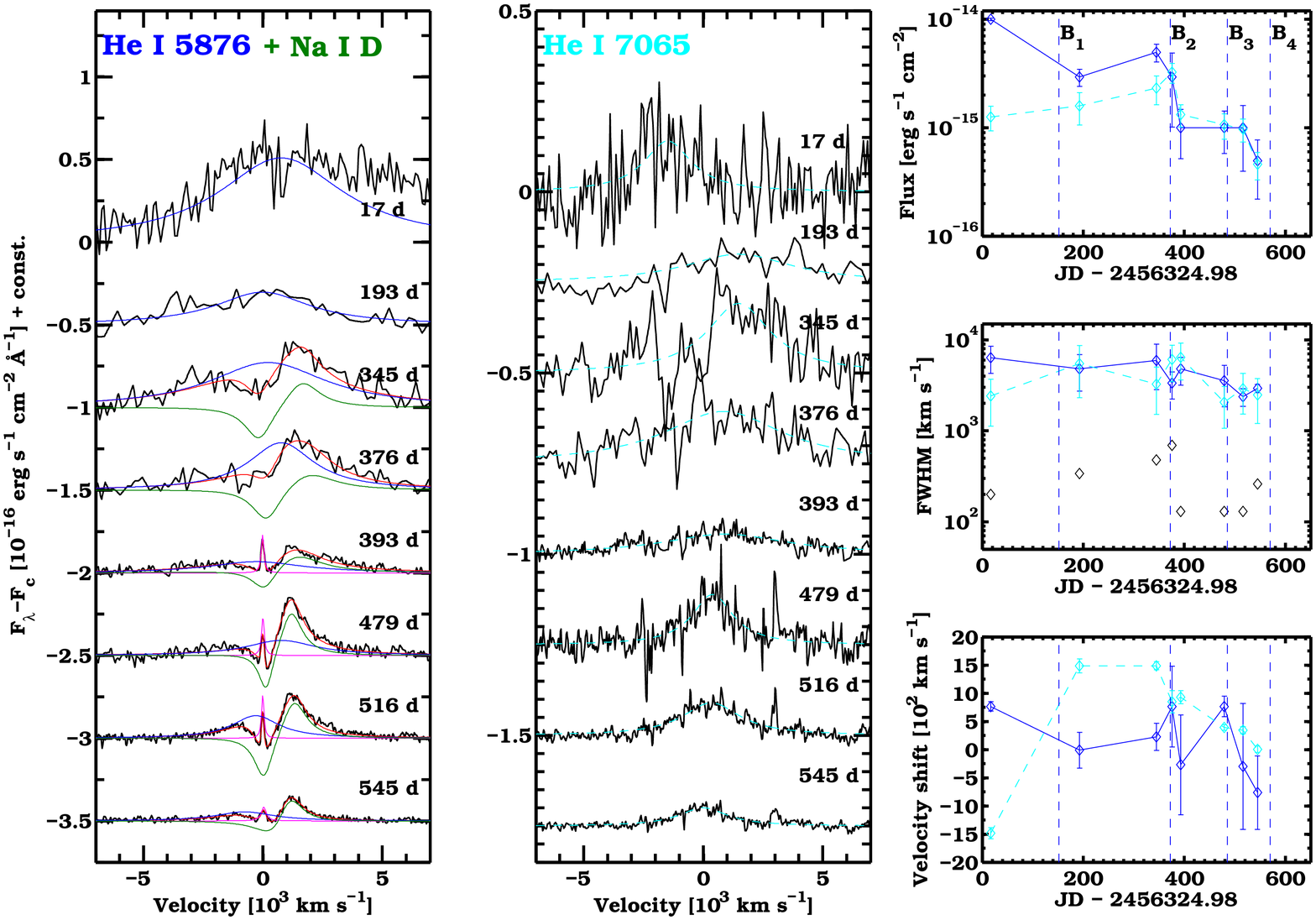}$
    \caption{Multi-component fits to the \ion{He}{I} and \ion{Na}{I} D lines of iPTF13z, with the fluxes, FWHM velocities, and velocity shifts plotted in the panels to the right. The black diamonds  in the middle right plot show the resolution of the spectra. The epochs of the bumps are marked with vertical dashed blue lines (labelled B$_{1-4}$,  upper right panel). The blue symbols and lines in the subplots to the right show measurements of the \ion{He}{I} $\lambda5876$ line; cyan shows measurements of the \ion{He}{I} $\lambda7065$ line. From 345 days  onwards, a double Lorentzian (in green) was also fitted to account for the \ion{Na}{i}~D line. A Lorentzian shown in magenta was fitted to the host galaxy \ion{He}{I}~$\lambda$5876.  Errors for the \ion{He}{I} lines were estimated as in Fig. \ref{fig:13zHalpha} for H$\alpha$.}
    \label{fig:13zHeI}
\end{figure*}
 
\subsubsection{The \ion{He}{I} emission lines\label{sec:He}}
The analysis of the H$\alpha$ line was repeated for the \ion{He}{I} emission lines, as shown in Fig.~\ref{fig:13zHeI}. The \ion{He}{I}~$\lambda$7065 line was fit with a single Lorentzian component (cyan symbols and dashed lines) as no narrow component was clearly detected at any epoch, from  host or SN.

The \ion{He}{I}~$\lambda$5876 line showed a single peak in the first two spectra (the first one is not perfectly matched by a Lorentzian, see the blue line), albeit with some asymmetry:  the red side is brighter than the blue side. In the spectra from 345 and 376 days, a double peak due to the presence of a \ion{Na}{i}~D P Cygni profile on top of the broad He emission (again blue line) was detected, and we disentangled the two profiles by fitting two additional Lorentzians to the \ion{Na}{i}~D line (the result is the green P Cygni profile in Fig. \ref{fig:13zHeI}). Starting from 393 days, a narrow \ion{He}{i}~$\lambda$5876 component at zero velocity (mainly due to galaxy emission) was clearly detected, and fit with another Lorentzian to remove its contamination (magenta line).

The \ion{He}{i} emission from the SN was thus characterised by relatively broad lines, whose FWHM slowly decreased from $\sim$6000 km~s$^{-1}$ to $\sim$2000 km~s$^{-1}$. This resembled the evolution of the broad component of H$\alpha$. The line flux of \ion{He}{i}~$\lambda$5876 decreased following the light curve, with a secondary peak in correspondence with bump B$_2$. The \ion{He}{i}~$\lambda$7065 line was fainter than the $\lambda$5876 line, and had a similar trend in flux, with the exception of the first spectrum where the line was relatively fainter. This line was also strongly blueshifted at 17~days, whereas it became redshifted at later epochs. The redshift decreased with time until it disappeared at 545~days. The velocity and the evolution of this redshift was similar to that observed for H$\alpha$. The redshift measured for the \ion{He}{I}~$\lambda$5876 line is more problematic to assess due to the complex \ion{Na}{i}~D contamination.

\subsubsection{\ion{Ca}{II}, \ion{Fe}{II}, and P Cygni profiles\label{sec:vel}}

The \ion{Ca}{II} near-IR triplet at $\lambda\lambda8498, 8542, 8662$ became clearly distinguishable from the surrounding continuum at 393~days (see Fig.~\ref{fig:spec13z}). It was discernible in the spectra from 345 and 376 days, but their lower resolution and lower S/N made it less conspicuous. At 393 days, the \ion{Ca}{II} triplet displayed narrow P Cygni profiles. The \ion{Ca}{II} K $\lambda 3934$ and \ion{Ca}{II} H $\lambda 3968$ lines were visible in the spectrum from 376 days, and possibly detected at 545~days (Fig. \ref{fig:spec13z}). There was no coverage at such short wavelengths in the other spectra at $>$ 300 days. A comb of \ion{Fe}{II} lines appeared around 4900 -- 5400~\AA\ from 345 days (Fig.~\ref{fig:spec13z}). We identify \ion{Fe}{II} $\lambda\lambda$4924, 5018, 5169, 5197, 5284, 5316. Owing to these lines, this region of the spectrum displays a sawtooth shape with a series of P Cygni profiles. 

From Fig.~\ref{fig:spec13z}, it is clear that H$\beta$ exhibited a relatively broad P Cygni profile which was contaminated by a strong narrow component in emission. This profile was observed from at least 376 days. At that epoch it was also observed in H$\gamma$ and H$\delta$.

\begin{figure*}
   \centering
    \includegraphics[width=16cm,angle=0]{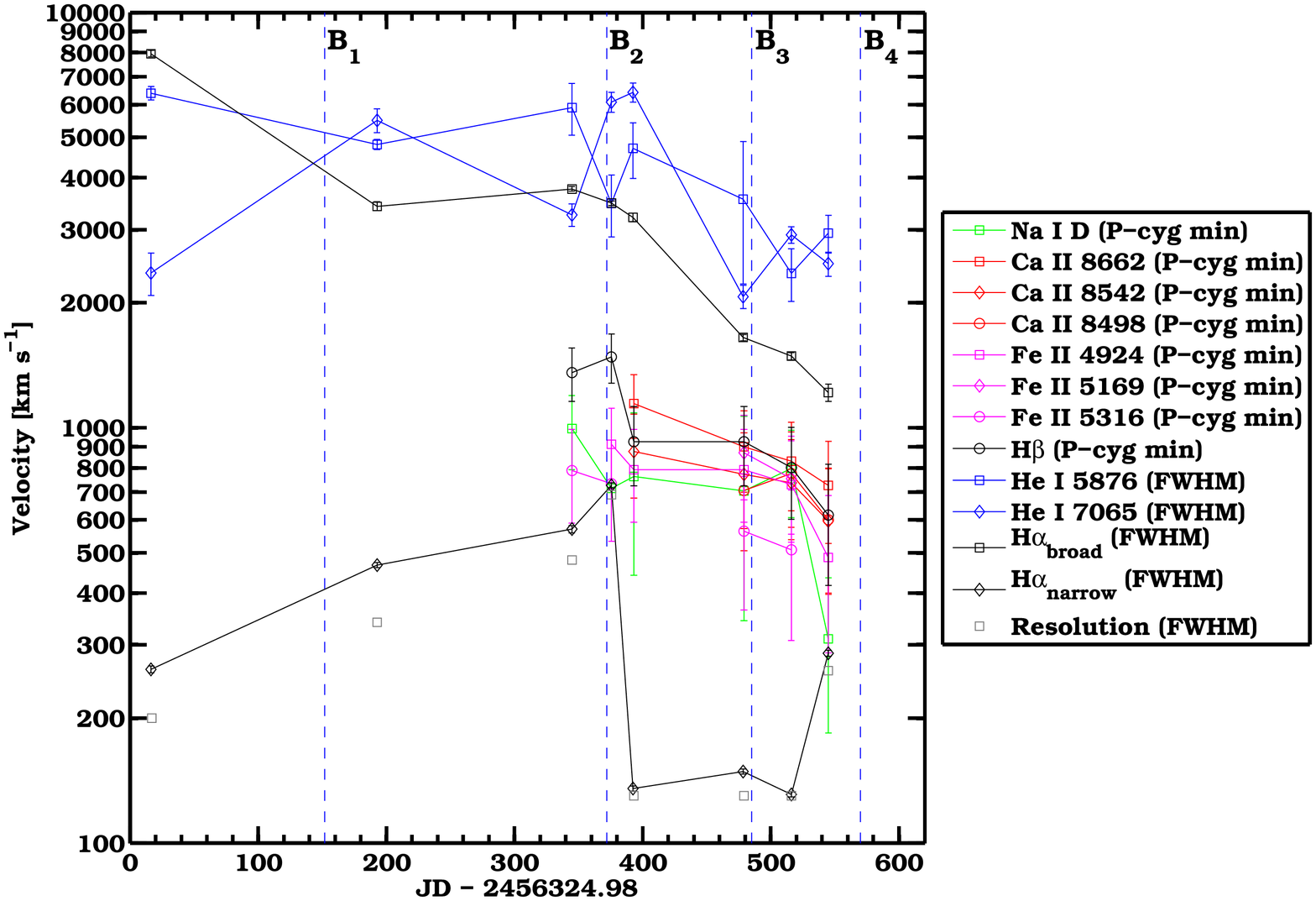}
    \caption{Velocities of iPTF13z as measured from the P Cygni absorption minima of the \ion{Na}{i}~D, \ion{Ca}{II}, and \ion{Fe}{II} profiles, and from the FWHM of the H$\alpha$ and \ion{He}{I} emission lines. Vertical blue lines show the epochs of the light curve bumps, labelled B$_{1-4}$.}
    \label{fig:velocity}
\end{figure*}

In Fig.~\ref{fig:velocity} we summarise the velocity measurements of the spectral lines, especially from the P Cygni absorption minima. At the epochs where we started seeing the \ion{Ca}{II} triplet, its P Cygni minimum velocity was $\sim 1000~\rm km ~s^{-1}$, and it slowly decreased to $\sim 700~\rm km ~s^{-1}$. The velocity was low enough to not cause blending of the individual lines (Fig.~\ref{fig:velocity}). The velocities from the minima of the \ion{Fe}{II} P Cygni profiles were  similar, as were those of H$\beta$.  The \ion{Na}{i}~D P Cygni absorption velocity is also shown, as derived from the fit in Fig.~\ref{fig:13zHeI}, and they closely match the other P Cygni line measurements.

The FWHM of the broad components of both \ion{He}{i} lines and H$\alpha$ were a few thousand km~s$^{-1}$, whereas the unresolved narrow component of H$\alpha$ traces the spectral resolution and was as low as $\lesssim$130 km~s$^{-1}$.

The appearance of the \ion{Ca}{II} near-IR triplet in iPTF13z was similar to that in SN 2005cl at 65 days after $R$-band maximum \citep{kiewe12}, SN 2005kj at 88 days after discovery \citep{taddia13}, and SN 2011ht at 125 days after discovery \citep{mauerhan2013b}. These three SNe all had low expansion velocities, so the triplet was not blended. A comb of \ion{Fe}{II} lines resembling that of iPTF13z was seen in SN 2005cl at 65 days after $R$-band maximum \citep{kiewe12}, in \object{SN 1995G} at 36 days after discovery \citep{pastorello02}, and in SN 1994W at 57 days after explosion \citep{sollerman98}.

\subsection{Spectral lines and SN velocities\label{sec:lineregion}} 
To understand the nature of iPTF13z, we need to discuss the emission mechanism and the location of line formation for the spectral lines presented in Sect.~\ref{sec:specanalysis}. 

The broad emission line components of H$\alpha$ and \ion{He}{i} showed similar velocity values and evolution. Actually H$\alpha$ appeared slightly slower than \ion{He}{i} at most epochs, even though this difference might have been due to the contamination of [\ion{N}{ii}] and \ion{He}{i}~$\lambda$6678 lines, and to the \ion{Na}{i}~D contaminating \ion{He}{i}~$\lambda$5876. The broad component could be interpreted as the velocity of the ionised ejecta. Alternatively, the broadness of the line might come from Thomson scattering in a thick CSM \citep{chugai01}.

It is also possible that the typical ejecta velocities are represented by the P Cygni minimum velocities of the \ion{Fe}{ii} and the other metal lines presented in Sect.~\ref{sec:vel}. These lines were detected not earlier than one year after discovery when parts of the CSM  had possibly become optically thin, allowing us to observe through it, while most of the CSM was still thick and kept scattering the prominent H$\alpha$ and \ion{He}{i} emission lines. The CSM velocity can then be represented by the FWHM of the narrow H$\alpha$ component, which was $\lesssim$130 km~s$^{-1}$ (the lines are not fully resolved).

\subsection{ Host galaxy\label{sec:hostgal}}
iPTF13z was found in the irregular galaxy SDSS J160200.05+211442.3 (alternative name \object{SDSSCGB 53789.4}). \citet{mcconnachie09} gives the position of the galaxy as $\alpha=$ $\rm 16^h$$\rm 02^m$$0\fs1$, $\delta =$ $21\degr 14\arcmin 42\farcs4$ (J2000.0). From SDSS \citep[Data Release 9;][]{ahn12}, we obtained magnitude $r = 18.96\pm0.02$ and colours $(u-g) = 1.19$ and $(g-r) = 0.12$ for the galaxy. This places the galaxy in the blue portion of the galaxy colour-colour diagram of \citet{strateva01}.

For SDSS J160200.05+211442.3 (hereafter J1602), the absolute $r-$band magnitude is $M_r = -16.99$. The galaxy is thus a dwarf whose absolute magnitude is comparable to that     of the Small Magellanic Cloud (SMC, with $M_{V}^{SMC} = -16.7$; using magnitude and distance modulus from NED). The apparent dimensions of J1602 (from SDSS DR 6, via NED) corresponds to $2.7 \times 1.7$ kpc. This is somewhat smaller than the dimensions of the SMC on the plane of the sky, which is around $4\times2$ kpc \citep[see e.g.][]{jac16}. The SDSS image of J1602 (Fig. \ref{fig:finder13z} inset) shows an elongated, slightly bent shape. The physical size and irregular appearance of this dwarf galaxy is also  comparable to the SMC.

To estimate the star formation rate (SFR) of J1602, we used the relation given by \citet[][eq. 3]{rosa02} between the galaxy SFR and luminosity of the \ion{[O}{II]} $\lambda 3727$ line. We measured this \ion{[O}{II]} line flux in our flux calibrated spectrum from 545 days and obtained an SFR of $0.046 ~ M_{\sun} ~yr^{-1}$. The SFR in the Magellanic clouds is $\sim 0.1 ~ M_{\sun} ~yr^{-1}$ \citep{harris04, harris09}. In summary, the host of iPTF13z is a blue irregular star-forming dwarf galaxy. The two iPTF13z-like events SN 2010mc and LSQ13zm also took place in dwarf galaxies \citep{ofek13a,tartaglia16}.

We used the spectroscopic method by \citet{pettini04} to estimate the metallicitiy of J1602. In the latest spectrum available to us, from Keck 2 at 958 days, the [N II] $\lambda6583$ line is clearly visible and gives $ N2~\equiv~\log\{[\ion{N}{II}] ~\lambda6583~/~H\alpha\}~=~-1.2138$. With this value, \citet*[eq. 2]{pettini04} gives us $\log{[O/H]} + 12 = 8.19 \pm 0.18$. The metallicities of LMC and SMC are $\log{[O/H]}_{LMC} + 12 = 8.37 \pm 0.22$ and $\log{[O/H]}_{SMC}~+~12~=~8.13~\pm~0.10$, respectively \citep{russell90}. The host of iPTF13z can therefore be considered an analogue of the SMC also in terms of metallicity, with a clear subsolar metallicity.
 
iPTF13z, which is a long-lasting SN~IIn (Fig. \ref{fig:complc}), shares the low metallicity \citep{taddia15} of its location with this subclass of SNe~IIn, in contrast to fast declining SNe IIn similar to SN 1998S which tend to explode at higher ($\sim$ solar) metallicity.

\subsection{A model of iPTF13z\label{sec:model}}
In this section we propose a simple model for the light curve of iPTF13z, which allows us to estimate the density of the CSM and the mass-loss rate of the SN progenitor.

\subsubsection{Explosion epoch estimate\label{sec:explodate}}
In order to model the SN, we need to constrain the explosion epoch. At the time of discovery, iPTF13z was emerging into the morning sky after conjunction with the Sun. The last non-detection of iPTF13z (from CRTS) was 129 days before the time of discovery. This gives weak observational constraints on when the SN exploded. 
 
One way to estimate the time of explosion is to compare the earliest part of the iPTF13z light curve to other SNe IIn having well-constrained explosion epochs and similar early light curve shapes and luminosities. For this comparison, the literature was searched for SNe IIn with published light curves in the $r/R$ band having pre-maximum photometry. The slope of the post-peak light curve decline was estimated by fitting a low-order polynomial to the light curves of each SN. The absolute magnitudes of the SNe were estimated using cosmology parameters from Sect. \ref{sec:distext} and MW extinction from \citet{schlafly11}. iPTF13z had an initial ($0 - 90$ days) linear slope of 0.016 mag day$\rm^{-1}$. Among the slopes computed for the literature sample, two SNe stand out as analogues to the early iPTF13z: PTF10aazn (= \object{SN 2010jj}) and \object{iPTF13agz}. We note that  SN 2010jj and iPTF13agz are good iPTF13z analogues in the $R$ band; however, inspection of their spectra reveals that there also are dissimilarities between them and iPTF13z.

The slope for SN 2010jj in the $R$-band observations was found to be 0.015 mag day$\rm^{-1}$. Using a suitable offset in time, a good match to the light curve of iPTF13z was obtained (the magnitudes were not shifted as they already matched on the decline). \citet{ofek14outb} gives the explosion epoch of SN 2010jj as MJD 55512 ($1.3$ days before the start of its light curve). Translated into the time scale of iPTF13z, this gives an estimate of the explosion date of iPTF13z at 26 days before discovery. Applying the same technique to the $R$-band light curve of iPTF13agz, including a brightening of its light curve by 0.4 mag, we also obtained a good fit to the early iPTF13z light curve. Using the iPTF13agz explosion epoch \citep{ofek14rise} gives an estimated explosion date of iPTF13z at 56 days before discovery. The mean and standard deviation of the two estimates gives JD~$2456284\pm21$ as an estimate for the explosion epoch of iPTF13z, suggesting that the SN was about $40\pm21$ days old at the time of discovery. In the  rise-time analysis by \citet{ofek14rise}, SNe IIn \object{PTF10tyd} and \object{PTF12ksy} reach peak brightnesses comparable to our lower limit for iPTF13z and have rise times $\approx 20$ days, suggesting our rise-time estimate for iPTF13z to be reasonable. In Sect. \ref{sec:CSMmod} we initially assume that iPTF13z was discovered 40 days after explosion. As the explosion epoch is weakly constrained, we  also evaluate how the weak constraint on explosion epoch affects the CSM properties computed in Sect. \ref{sec:CSMmod}.

\subsubsection{CSM interaction model\label{sec:CSMmod}}
Using our explosion epoch estimate, we proceed to estimate the CSM density as a function of radius and the mass-loss rate in the decades before the SN exploded.

\begin{figure}[h!]
   \centering
    \includegraphics[width=9cm]{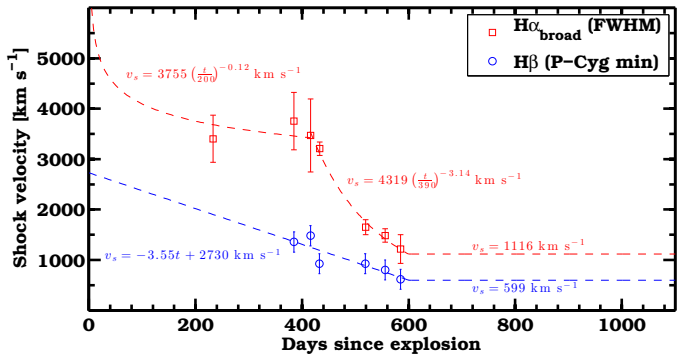}
    \caption{Shock velocity of iPTF13z assuming it is traced either by the H$\beta$ P Cygni absorption or by the H$\alpha$ broad FWHM. Explosion is assumed to occur 40~days before discovery. Equations of the fits are given in corresponding colours, for time $t$ given in days. The fits are used to describe the time dependence of the shock velocity, using our spectra.}
    \label{fig:velfit}
\end{figure}

For a light curve powered by CSM interaction, the bolometric luminosity can be expressed as $L = 2 \pi \epsilon \rho_{CSM} v_{s}^3 r_{s}^2$ (c.f. e.g. \citealt[eq. 1]{chugai91} and \citealt{salamanca98}), where $v_{s}$ is the shock velocity, $r_{s}$ the shock radius, $\rho_{CSM}$ is the density of the CSM, and $\epsilon$ is the conversion efficiency from kinetic energy to radiated energy. This model assumes a spherically symmetric SN, which is presumably  a simplification. If luminosity and shock velocity are given, we can estimate the CSM density as

\begin{equation}
\label{eq:vel}
\rho_{CSM} = \frac{L}{2 \pi \epsilon v_{s}^3 r_{s}^2},
\end{equation}

with the shock radius expressed as $r_s = \int_0^t{v_s dt'}$. Below, we use piecewise fits (Fig. \ref{fig:velfit}) to velocities (obtained from our spectra) to calculate values of $r_s$. We set the conversion efficiency as $\epsilon = 0.3$, following \citet{moriya14}. We calculate the bolometric luminosity by converting the absolute $r$-band light curve using Eq. \ref{eq:lum}. Bolometric corrections are ignored here, as in Sect. \ref{sec:colour}.

Based on our discussion in Sect.~\ref{sec:lineregion}, we first consider the P Cygni velocity of H$\beta$ as a proxy for the shock velocity $v_{s}$. We fit the observed H$\beta$ velocities (blue symbols in Fig.~\ref{fig:velfit}) with a linear function, to extrapolate its values at epochs earlier than 385 days since explosion. At epochs later than 600 days after explosion, where spectra are lacking, we assume that the velocity is constant and equal to the velocity at 600 days. The best fit is shown with a blue dashed line in Fig.~\ref{fig:velfit}.

\begin{figure}
   \centering
    \includegraphics[width=9cm]{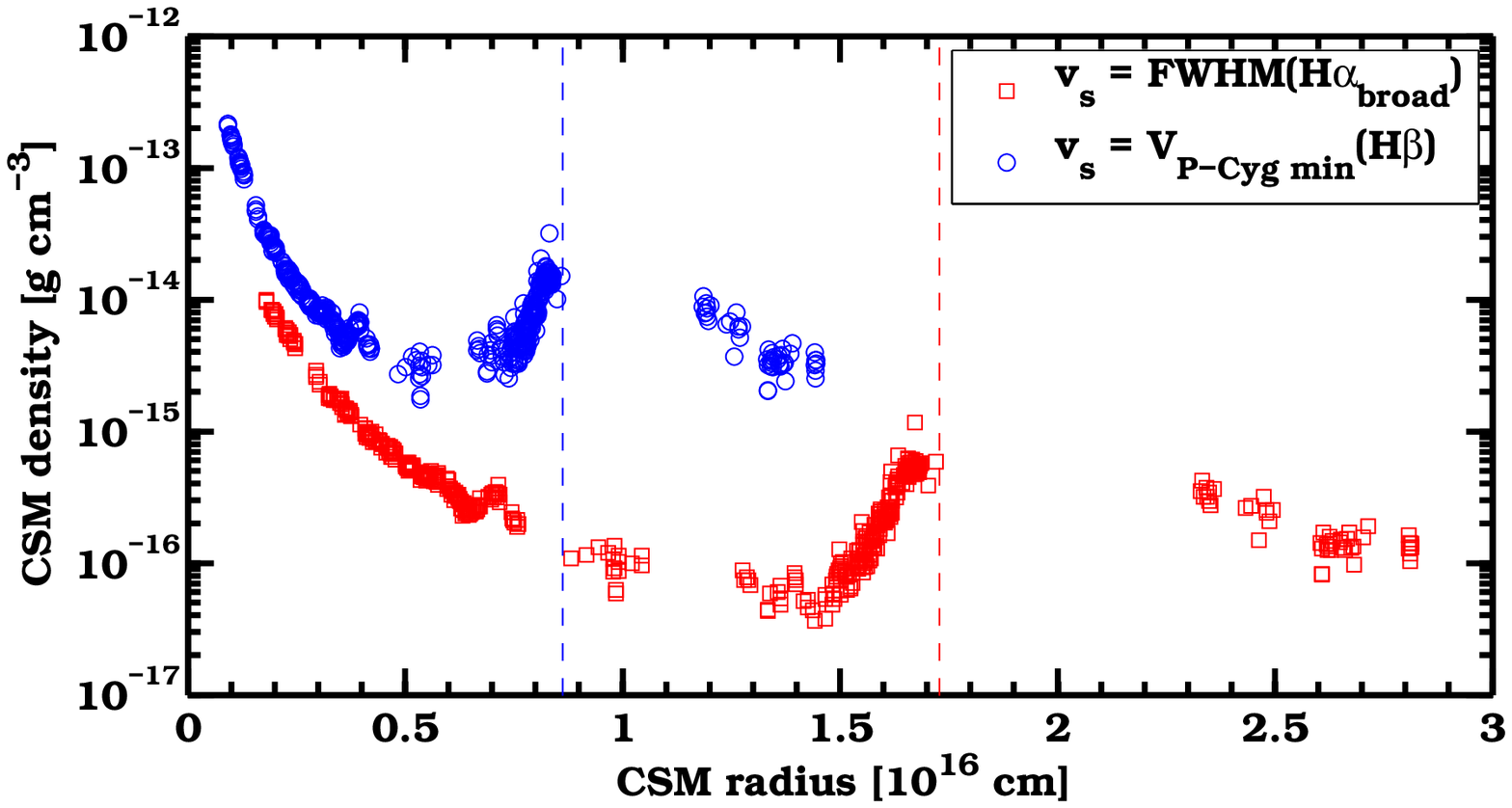}\\
    \includegraphics[width=9cm]{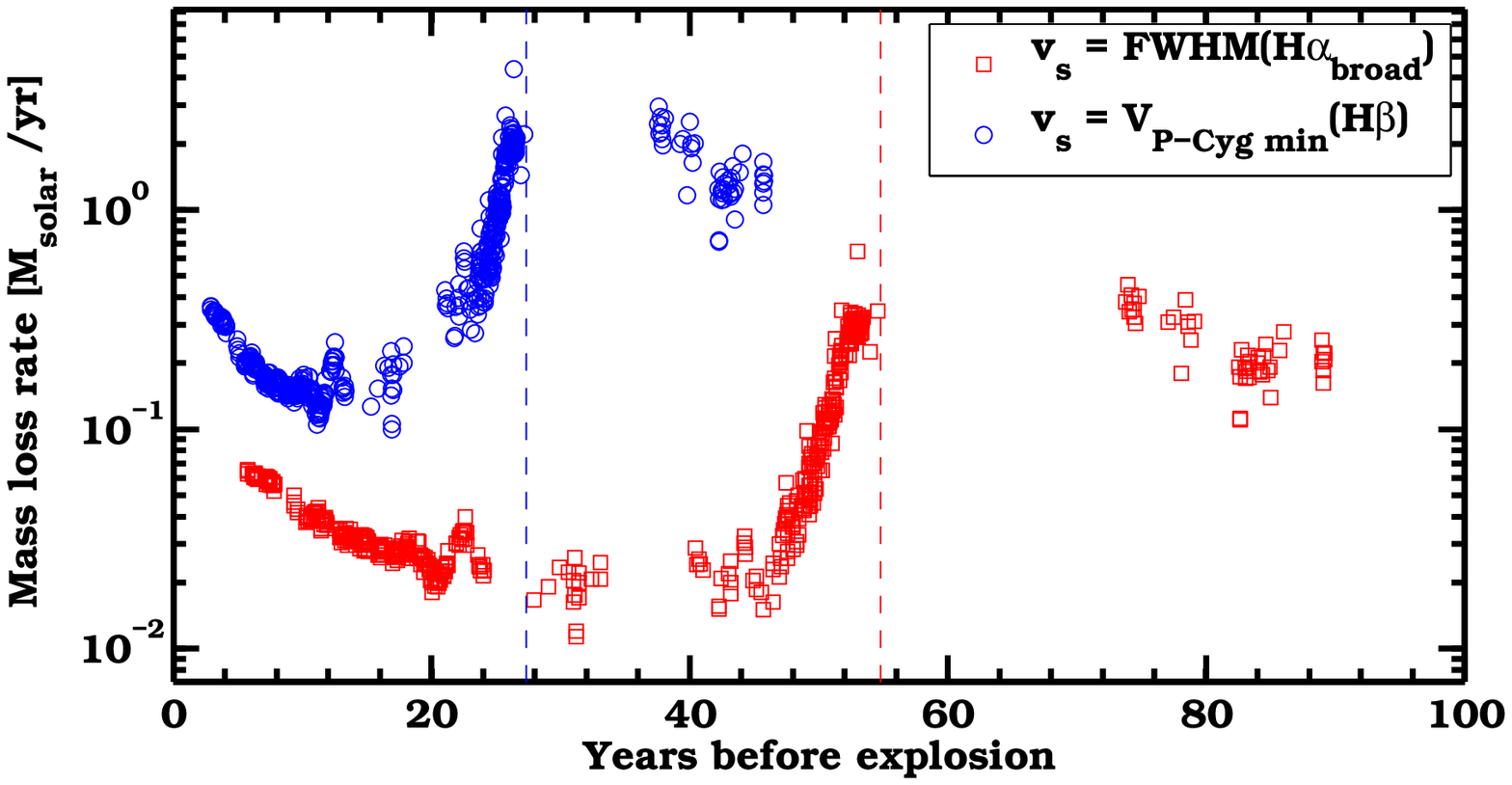}
\caption{CSM density structure (top panel) and mass-loss rate (bottom panel), as estimated from the bolometric light curve of iPTF13z with the model described in Sect.~\ref{sec:model}. For these plots, we use $\epsilon = 0.3$, $v_w = 100\rm ~km ~s^{-1}$, and SN shock velocity, as plotted in Fig. \ref{fig:velfit}. Dashed vertical lines show, in corresponding colour,  where the assumption about constant velocity later than 600 days is used in the CSM density and mass-loss calculations.}
\label{fig:CSM}
\end{figure}

With the estimated luminosity and velocity, we use Eq. \ref{eq:vel} to derive the CSM density structure. The result based on H$\beta$ P Cygni absorption is shown in blue in Fig. \ref{fig:CSM} (top panel). High CSM densities between 10$^{-15}$ and 10$^{-13}$ g~cm$^{-3}$ are obtained, with peaks corresponding to the light curve bumps.

The CSM density can be converted into a mass-loss rate (assuming constant wind velocity, $v_w$, for simplicity), using the  equation

\begin{equation}
\dot{M} = 4 \pi r^2 \rho_{CSM} v_w
\end{equation}

(see \citealt[][eq. 12,]{imshennik74} and \citealt{moriya13a}). As $r = v_w t$, we can estimate the mass-loss history of the progenitor star, i.e. $\dot{M}(t)$. The mass-loss history is shown in the bottom panel of Fig. \ref{fig:CSM}, where we assume $v_w = 100\rm ~km ~s^{-1}$. This is of the order of the partially resolved FWHM of the narrow H$\alpha$ component (Sects.~\ref{sec:Halpha} and ~\ref{sec:lineregion}). Very large mass-loss rates of $\approx 0.1-2 ~M_{\sun} ~yr^{-1}$ are found. In correspondence with the light curve bumps, the mass-loss rate of iPTF13z shows local maxima. These are consistent with eruptions that occurred between 10 and 50 years before the explosion. This model shows that the eruptions producing the bumps observed in the light curve occurred years before we imaged that portion of the sky with iPTF.

We also observed a recent pre-discovery outburst (around $-$210 days) confirming that the progenitor of iPTF13z underwent eruptive episodes until the final year before core-collapse. Could we have observed the interaction between this outburst material and the SN ejecta? Assuming that iPTF13z was 40 days old at the time of discovery means that the pre-discovery outburst we saw took place around 170 days before explosion. Assuming ejecta velocity 2000 $\rm km~s^{-1}$ (Fig. \ref{fig:velocity}) and outburst velocity $\sim$130 $\rm km~s^{-1}$ (based on the width of the narrow H$\alpha$ line), \citet[][eq. 1]{graham14} gives $-28$ days as the epoch of interaction between the $-$210 days outburst material and the SN ejecta. Any interaction between material from the pre-discovery outburst at $-$210 days and the SN ejecta thus took place before our discovery of the SN.

If we perform our modelling adopting the shock velocity from the FWHM of the broad H$\alpha$ component (fit with three functions depending on the SN age, as shown with a dashed red line in Fig.~\ref{fig:velfit}), the result is CSM density in the range 10$^{-16}$ to 10$^{-14}$ g~cm$^{-3}$ and mass-loss rates in the range $0.01-0.4$ $~M_{\sun}~yr^{-1}$ (red symbols in Fig.~\ref{fig:CSM}). Integrating the mass-loss rate over the time interval that we can explore before explosion, the total CSM mass ejected in the 90 years before explosion would be $\sim$13 $~M_{\sun}$ for this model.

If we instead use the model based on velocities from H$\beta$ P Cygni absorption, integration of the mass-loss rate over the time we can explore before explosion ($\sim$50 years), the found total CSM mass ejected is $\sim 48 ~ M_{\sun}$. To make significant light curve bumps, the mass of the CSM needs to be of the same order of magnitude as the SN ejecta mass. If the exploding star were an LBV (some of which have $\gtrsim 50 ~ M_{\sun}$ zero-age main sequence mass)  this would not be unphysical but less realistic.

We recall from Sect. \ref{sec:lineregion} that the broad component of H$\alpha$ emission and the H$\beta$ P Cygni absorption component represent the SN expansion velocity evolution in different ways. The different diagnostics will therefore lead to different results in our model.

Within the realm of our model, for a given shock velocity profile, the age of the SN at time of discovery does not strongly affect the CSM parameters obtained from the model. Varying the age at time of discovery between 1 day and 128 days, with the other parameters unchanged, affects the integrated CSM mass at the order of $\sim 10$ \%.

To further evaluate the results from our CSM model, we did a simple Monte Carlo experiment concerning the CSM mass. We generated 10000 sets of model parameters, with values drawn randomly from uniform distributions in the ranges $0.1 < \epsilon < 0.5$ and $40 < v_{w} < 130$ $\rm ~km ~s^{-1}$. Because of  the comparatively small impact of the age of the SN at discovery, we kept this at 40 days. Since uncertainty in the representation of the shock velocity is important, for each set of model parameters we generated velocity functions based on the H$\alpha$ and H$\beta$ lines (as in Fig. \ref{fig:velfit}) with a random increment or decrement (from ranges $\pm 200 \rm ~km ~s^{-1}$ for both lines) added to the velocities used for making the fits.

For both velocities, our set of model parameters gave a right tailed distribution of CSM masses. For our H$\alpha$ (FWHM) velocities, 68 \% of the Monte Carlo realisations gave CSM mass values in the range $7-20 ~M_{\sun}$. For the H$\beta$ (P Cygni) velocities, the range was $18-65 ~M_{\sun}$. These ranges can be seen to indicate the errors of our estimates, within the given model. The estimates of CSM and mass-loss properties should in any case be seen as order-of-magnitude estimates, as our assumptions of constant wind velocity and spherical geometry presumably are simplifications \citep[e.g.][]{dwarkadas11, moriya14massloss}. From the discussion of spectra in Sect. \ref{sec:lineregion}, we see that the SN and its CSM is most likely not spherically symmetric. The H$\alpha$ based CSM properties from Fig. \ref{fig:CSM}, rounded to the closest order of magnitude, suggest a CSM mass of $\sim 10 ~ M_{\sun}$, a CSM density of at least 10$^{-16}$ g~cm$^{-3}$ (higher in the shells), and a mass-loss rate of $0.01 ~M_{\sun}~yr^{-1}$ (higher during progenitor outbursts).

Our assumption that the light curve is powered by ejecta--CSM interaction is reasonable, as the spectral sequence of iPTF13z (Fig. \ref{fig:spec13z}) consistently  shows Balmer lines with narrow central components. In addition to the durability of the light curve (Fig. \ref{fig:lc13z}) and the brighter portions of the bumps ($\Delta m > 0$; Sect. \ref{sec:LC}), another sign of ejecta-CSM interaction is also shown by the light curve.

iPTF13z seems to become fainter before the brightness increase of a bump starts. This is evident for bump B$_1$, but also possible for bumps B$_2$, B$_3$, and B$_5$ in the decline-subtracted $r$-band light curve (Fig. \ref{fig:13zanalysis}b). To explain a light curve dip followed by brightening in the superluminous \object{SN 2006oz} \citep{leloudas12}, a model was proposed by \citet{moriya12}: SN ejecta collides with a dense CSM shell, producing energetic photons at the forward shock. This photoionises the CSM and the increased Thomson scattering in the CSM increases the opacity of the shell. The CSM shell thus blocks light from the region inside it, making a dip in the SN light curve. We note the similarity between the dips before iPTF13z bumps B$_1$, B$_2$, B$_3$, and B$_5$ and the dip modelled by \citet{moriya12}.

\section{Discussion\label{sec:discussprog}}
Our analysis and model provide some constraints on the nature of the progenitor star of iPTF13z.

The velocity of the un-shocked CSM can be deduced from the FWHM of the narrow partially resolved H$\alpha$ emission. This is of the order of 130 km~s$^{-1}$ or lower. This points to a SN progenitor star with a wind velocity that presumably corresponds to that of an LBV (which is given in the range 50 to 500 km~s$^{-1}$; see e.g. \citealp{dwarkadas11, kiewe12}). We observe multiple velocity components in the spectra and we simultaneously see emission lines that are electron scattered by a thick CSM, as well as lines with P Cygni profiles from expanding ejecta. Luminous blue variables like $\eta$ Car are asymmetric, allowing us to observe both optically thick and optically thin regions. By analogy, this suggests asymmetry in the CSM structure of iPTF13z.

The mass-loss rate inferred from the  modelling for the final decades of the progenitor star is variable in time and characterised by very high values ($\gtrsim$ 0.01 $M_{\sun}~yr^{-1}$ even during relatively quiet episodes in the final decades of the progenitor star). This gave rise to a CSM with large variations in density, which in turn produced the bumps in the light curve once it was shocked by the interaction with the SN ejecta. These aspects are most easily explained by eruptive events forming a large (radius $>10^{16}$~cm) and massive ($\sim 10$ $~M_{\sun}$) CSM surrounding the SN. Again, these results are indicative of an LBV progenitor affected by large mass ejections \citep[see e.g.][]{smith14}.

In the case of iPTF13z, we directly observed an outburst prior to the main explosion. The average magnitude of the outburst was $M_r \approx -15$, which can be compared to the LBV light curves compiled by \citet[][Figs. 7 and 8]{smith11_lbv}. A typical magnitude range for LBV outbursts is $-14 < M < -12$. For example, the year 1843 outburst of $\eta$ Car reached $ M \approx -14$ mag at its peak. The comparable absolute magnitude makes the 2012 pre-discovery event of iPTF13z a candidate for a luminous LBV outburst. We recall that SN IIn 2005gl \citep{galyam07, galyam09} has observational indications of possibly having an LBV star as its progenitor. Examples of interacting CC SNe which have shown indications of distinct shell-like structures in their CSM are  SN 1996L \citep{benetti99} and the transitional Type Ib/IIn \object{SN 2014C} \citep{milisavljevic15}. Such CSM shells could have arisen from eruptive episodes of the progenitor stars shortly before they exploded.

There is the possibility that our suggested LBV progenitor star was in a binary system. The CSM formed by the large mass ejections from the progenitor star might be stirred into a spiral shape by a companion star. Then, when the SN explodes, its collision with the denser parts of the spiral could give rise to a series of bumps in the LC. In the radio light curve of Type II SN 1979C, \citet{schwarz96} interpreted the modulations in the light curve as being due to the presence of a companion modulating the CSM density into a spiral. The well-studied LBV $\eta$ Car is known to be in a binary system \citep[e.g.][]{daminelli96, madura12} and has a complex and irregular CSM. \citet{kotak06} proposed that mass loss from an LBV eventually exploding as a SN could generate a SN radio light curve with undulations (see also \citealp{moriya13b}). Late bumps were seen in radio light curves of Type IIb SNe 2001ig \citep{kotak06, ryder04} and 2003bg \citep{soderberg06}.

Red supergiants (RSGs) have also been discussed as progenitors of Type IIn SNe (e.g. \citealp{smith09_rsg}). \citet{fransson02}, studying \object{SN 1995N}, suggest that RSGs with strong stellar winds are SN~IIn progenitors. \citet{mackey14} proposed that external radiation (from a companion) can phototionise the stellar wind of a RSG, which in turn can confine CSM close to the star. Thus confined, the CSM will allow the RSG to give rise to a SN~IIn when exploding. Modelling by \citet{yoon10} point to RSGs with pulsationally driven superwinds as possible SN Type IIn progenitors. The upper limit estimated for the wind velocity of the iPTF13z progenitor ($\lesssim$130 km~s$^{-1}$) does not rule out a RSG as the progenitor. However, comparing typical mass-loss rates of RSGs ($\sim ~10^{-6} ~M_{\sun} ~yr^{-1}$; e.g. \citealp{mauron11}) to the values found from iPTF13z ($\sim ~10^{-2} ~M_{\sun} ~yr^{-1}$, with episodes of greater mass loss), this speaks against a red supergiant as the iPTF13z progenitor and points instead towards an LBV. The observed (LBV-like) pre-discovery outburst at $M_r \approx -15$ strengthens this indication that the progenitor of iPTF13z was an LBV.

We note that iPTF13z and the comparison events shown in Fig. \ref{fig:outburst_comp} have their precursor events clustering between absolute magnitude $-14$ and $-15$ and their main events (possibly CC SN explosions) clustering between absolute magnitude $-17$ and $-19$. This similarity was discussed by \citet{tartaglia16} for the precursors and main events of SNe 2009ip, 2010mc, and LSQ13zm, which are all included in Fig. \ref{fig:outburst_comp}. Whereas the precursor event and lower limit of the iPTF13z peak magnitude make it comparable to the other Fig. \ref{fig:outburst_comp} events, it is distinguished from them by its pronounced series of light curve bumps and by its SN 1988Z-like durability (Fig. \ref{fig:complc}).

It is hard to be conclusive about the finality of iPTF13z. The unresolved controversies about the nature of SNe 1961V \citep[e.g.][]{kochanek11, smith11_lbv, vandyk12} and 2009ip illustrate how hard it is to draw  conclusions about core-collapse, even for nearby ($\sim 10$~Mpc) Type IIn events. Comparing the brightness of iPTF13z (peak magnitude $M_r \lesssim -18.3$) with that of other SNe IIn places iPTF13z in the brighter portion of the range of conventional SN IIn maximum magnitudes (Sect. \ref{sec:intro}). Considering this, one simple assumption would be that iPTF13z was a CC SN, whose ejecta interacted with denser regions of the CSM (caused by eruptive mass-loss episodes similar to the episode around $-210$ days). Another mechanism which could possibly account for a precursor having eruptive mass-loss episodes and eventually give rise to a SN Type IIn is a pulsational pair-instability SN \citep[PPISN; e.g.][]{barkat67, woosley17}.
 
Continued photometric monitoring of iPTF13z is desirable, in order to search for any new re-brightenings. Imaging of the host galaxy at high angular resolution \citep[see][]{smith16} is of interest to examine the environment of iPTF13z. The successor of iPTF, the Zwicky Transient Facility \citep[ZTF;][]{spie14}, is expected to start operations on the P48 telescope in 2017. The ZTF and other upcoming transient surveys will allow continued photometrical monitoring of SNe IIn in order  to carry on the search for new pre-explosion outbursts and light curve bumps displayed by this heterogeneous SN type.

\begin{acknowledgements}
Thanks to John Telting, Auni Somero, Paul Vreeswijk, Ofer Yaron, Kunal Mooley, Ragnhild Lunnan, Danny Khazov, and Gina Duggan for making observations of iPTF13z. Thanks to Peter Lundqvist and Mattias Ergon for discussions. Thanks to Andrew Drake and Ashish Mahabal for assisting with access to the Catalina Real-time Transient Survey images. We gratefully acknowledge the support from the Knut and Alice Wallenberg Foundation. The Oskar Klein Centre is funded by the Swedish Research Council. This research used a computer bought with a grant from the Alva and Lennart Dahlmark Foundation. TJM is supported by the Grant-in-Aid for Research Activity Start-up of the Japan Society for the Promotion of Science (16H07413). EOO is the incumbent of the Arye Dissentshik career development chair and is grateful for support by grants from the Willner Family Leadership Institute Ilan Gluzman (Secaucus NJ), Israel Science Foundation, Minerva, Weizmann-UK, and the I-Core program by the Israeli Committee for Planning and Budgeting and the Israel Science Foundation (ISF). AGY is supported by the EU/FP7 via ERC grant No. 307260, the Quantum Universe I-Core program by the Israeli Committee for planning and funding, and the ISF, Minerva and ISF grants, WIS-UK ``making connections'', and Kimmel and YeS awards. The intermediate Palomar Transient Factory (iPTF) project is a scientific collaboration among the California Institute of Technology, Los Alamos National Laboratory, the University of Wisconsin, Milwaukee, the Oskar Klein Centre, the Weizmann Institute of Science, the TANGO Program of the University System of Taiwan, and the Kavli Institute for the Physics and Mathematics of the Universe. This research used resources of the National Energy Research Scientific Computing Center, a DOE Office of Science User Facility supported by the Office of Science of the U.S. Department of Energy under Contract No. DE-AC02-05CH11231. This work was supported by the GROWTH project funded by the National Science Foundation under Grant No. 1545949. This research used observations made with the Nordic Optical Telescope, operated by the Nordic Optical Telescope Scientific Association at the Observatorio del Roque de los Muchachos, La Palma, Spain, of the Instituto de Astrofisica de Canarias. This research used observations obtained with the Apache Point Observatory 3.5-meter telescope, which is owned and operated by the Astrophysical Research Consortium. Some of the data presented here were obtained at the W.M. Keck Observatory, which is operated as a scientific partnership among the California Institute of Technology, the University of California, and the National Aeronautics and Space Administration (NASA). The Observatory was made possible by the generous financial support of the W.M. Keck Foundation. The authors wish to recognise and acknowledge the very significant cultural role and reverence that the summit of Mauna Kea has always had within the indigenous Hawaiian community. We are most fortunate to have the opportunity to conduct observations from this mountain. This research used the SIMBAD and VizieR databases operated at CDS, Strasbourg, France. This research used the NASA/IPAC Extragalactic Database (NED) which is operated by the Jet Propulsion Laboratory, California Institute of Technology, under contract with NASA. This research used NASA's Astrophysics Data System.
\end{acknowledgements}

\bibliographystyle{aa}

\onecolumn
\clearpage

\begin{deluxetable}{cc}
\center
\tabletypesize{\scriptsize}
\tablewidth{0pt}
\tablecaption{\label{refstars13z} SDSS reference stars for P48 photometry of iPTF13z}
\tablehead{
\colhead{SDSS name} &
\colhead{AB magnitude} \\
\colhead{(DR9)} &
\colhead{($r$)}}
\startdata
J160147.79+211241.4 & 20.046 \\ 
J160148.39+211542.7 & 20.312 \\ 
J160148.46+211423.3 & 20.328 \\ 
J160152.08+211638.9 & 21.31 \\ 
J160152.66+211713.6 & 16.598 \\ 
J160152.74+211516.1 & 20.874 \\ 
J160153.87+211607.8 & 18.978 \\ 
J160154.84+211600.3 & 17.869 \\ 
J160156.41+211722.8 & 18.358 \\ 
J160157.02+211318.8 & 20.825 \\ 
J160157.84+211719.5 & 20.41 \\ 
J160158.28+211301.7 & 16.595 \\ 
J160200.58+211153.8 & 19.719 \\ 
J160202.05+211246.5 & 18.9 \\ 
J160202.24+211432.7 & 20.555 \\ 
J160202.57+211311.2 & 19.149 \\ 
J160203.04+211601.2 & 16.932 \\ 
J160203.21+211304.6 & 20.53 \\ 
J160205.22+211547.3 & 19.06 \\ 
J160205.77+211317.8 & 17.864 \\ 
J160206.00+211525.2 & 17.752 \\ 
J160206.20+211229.8 & 17.88 \\ 
J160207.60+211431.3 & 20.708 \\ 
J160208.31+211413.2 & 18.655 \\ 
J160210.67+211502.8 & 18.328 \\ 
J160211.26+211531.0 & 19.939 \\ 
J160211.66+211417.0 & 19.697 \\ 
J160212.32+211325.9 & 17.452 \\ 
J160212.52+211625.2 & 18.633
\enddata
\tablecomments{Reference stars for the P48 photometry. The SDSS DR9 names are based on J2000 equatorial coordinates of the stars. The $r$ magnitude, used when doing the relative photometry on iPTF13z, is taken from SDSS DR9.}
\end{deluxetable}

\begin{deluxetable}{ccc}
\center
\tabletypesize{\scriptsize}
\tablewidth{0pt}
\tablecolumns{12}
\tablecaption{\label{p48r13z_early} Early (0 $ - $166 days) P48 Mould $R$-band photometry of iPTF13z}
\tablehead{
\colhead{MJD} &
\colhead{Epoch} &
\colhead{$m_{R}$($\sigma_{R}$)}\\
\colhead{} &
\colhead{(days)} &
\colhead{}}
\startdata
56324.48 & 0.00 & 17.99(0.03) \\
56324.52 & 0.04 & 18.02(0.03)  \\
56324.56 & 0.08 & 17.98(0.03)  \\
56327.47 & 2.99 & 18.05(0.03)  \\
56327.51 & 3.03 & 18.05(0.03)  \\
56327.54 & 3.06 & 18.06(0.03)  \\
56328.47 & 3.99 & 18.10(0.03)  \\
56328.51 & 4.03 & 18.05(0.03)  \\
56328.55 & 4.07 & 18.06(0.02)  \\
56329.47 & 4.99 & 18.12(0.03)  \\
56329.50 & 5.02 & 18.08(0.03)  \\
56329.54 & 5.06 & 18.06(0.02)  \\
56330.46 & 5.98 & 18.11(0.03)  \\
56330.50 & 6.02 & 18.11(0.04) \\
56330.54 & 6.06 & 18.14(0.02) \\
56335.45 & 10.97 & 18.15(0.03) \\
... & ... & ...
\enddata
\tablecomments{AB magnitudes from PSF photometry (relative photometry based on SDSS) on host subtracted images. The middle column gives the epoch of the measurement relative to time of SN discovery ($\rm JD~ 2456324.98$). The right column gives $m_R$($\sigma_R$) with statistical error (1 $\sigma$) in parentheses. The full version of this table is only available via the CDS. The trimmed portion of the table shown here serves as a guide to its content.}
\end{deluxetable}

\onecolumn

\begin{deluxetable}{ccccccc}
\center
\tabletypesize{\scriptsize}
\tablewidth{0pt}
\tablecolumns{7}
\tablecaption{\label{13zp60P48} Late ($>$ 200 days) P48 and P60 photometry of iPTF13z}
\tablehead{
\colhead{MJD} &
\colhead{Epoch} &
\colhead{$m_B$($\sigma_B$)} &
\colhead{$m_{g}$($\sigma_{g}$)}&
\colhead{$m_{r}$($\sigma_{r}$)}&
\colhead{$m_{R}$($\sigma_{R}$)}&
\colhead{$m_{i}$($\sigma_{i}$)}\\
\colhead{} &
\colhead{(days)} &
\colhead{P60} &
\colhead{P60}&
\colhead{P60}&
\colhead{P48}&
\colhead{P60}}
\startdata
56528.21 & 203.73 & ... & 20.62(0.15) & 19.93(0.10) & ... & 19.95(0.10) \\ 
56539.25 & 214.77 & ... & ... & 19.80(0.06) & ... & 20.17(0.10) \\ 
56541.18 & 216.70 & ... & 20.87(0.08) & ... & ... & ... \\ 
56548.51 & 224.03 & ... & 20.92(0.17) & 19.60(0.06) & ... & 19.89(0.11) \\ 
56555.14 & 230.66 & ... & ... & 19.66(0.05) & ... & 19.92(0.08) \\ 
56556.19 & 231.71 & ... & 20.44(0.17) & ... & ... & ... \\ 
56559.46 & 234.98 & ... & ... & ... & 20.02(0.12) & ... \\ 
56560.62 & 236.14 & ... & ... & ... & 20.43(0.18) & ... \\ 
56561.19 & 236.71 & ... & ... & ... & 20.60(0.27) & ... \\ 
56563.14 & 238.66 & ... & ... & 19.97(0.09) & 20.00(0.12) & 20.17(0.09) \\
... & ... & ... & ... & ... & ... & ...
\enddata
\tablecomments{AB magnitudes from PSF photometry (relative photometry based on SDSS) on host subtracted images. The magnitudes are given as $m_{band}$($\sigma_{band}$), with the statistical error (1 $\sigma$) in parentheses. The epoch of the measurements are given relative to time of SN discovery ($\rm JD~ 2456324.98$). For the dates given in this table, measurements made within 1 day are considered to be simultaneous. The full version of this table is only available via the CDS. The trimmed portion of the table shown here serves as a guide to full content.}
\end{deluxetable}

\begin{deluxetable}{lcllcll}
\centering
\tabletypesize{\scriptsize}
\tablewidth{0pt}
\tablecaption{Spectral observations log for iPTF13z\label{speclog}}
\tablehead{
\colhead{UT Date} &
\colhead{Epoch (days)} &
\colhead{Telescope} &
\colhead{Instrument}&
\colhead{Resolution $\rm ~(km ~s^{-1})$}&
\colhead{Spectral range (\AA )}&
\colhead{Comments}}
\startdata
2013 February 18 & 17.08 & P200  &  DBSP & 200 & 3210-10100 &Classification spectrum\\    
2013 August 13 & 192.7 & APO  &  DIS  & 340 & 3340-8500 & \\
2014 January 12 & 344.79 & NOT  &  ALFOSC & 480 & 5000-9120 &  \\   
2014 February 12 & 375.74 & NOT  & ALFOSC & 690 & 3400-9120&  \\   
2014 March 1 & 393.11 & Keck 2  & DEIMOS  & 130 & 4900-10130&  \\         
2014 April 27 & 450.15 & Keck 2  & DEIMOS  & 130 & 4910-10150 & Noisy spectrum \\   
2014 May 26 & 479.05 & Keck 2  & DEIMOS  & 130 & 4920-10140&  \\  
2014 July 2 & 515.97 & Keck 2  & DEIMOS  & 130 & 4550-9600&  \\    
2014 July 31 & 544.87 & Keck 1 & LRIS &  260 & 3070-10300& \\ 
2015 January 26 & 723.76 & NOT & ALFOSC & 500 & 3650-9110& Dominated by host \\
2015 September 17 & 957.75 & Keck 2  & DEIMOS  & 140 & 4750-9600 & Dominated by host 
\enddata
\tablecomments{For our spectroscopy, we used the Double Spectrograph (DBSP) on the 5 m telescope (P200) at Palomar Observatory, the Dual Imaging spectrograph (DIS) on the Astrophysical Research Consortium 3.5 m telescope at Apache Point Observatory (APO; New Mexico, USA), the Andalucia Faint Object Spectrograph and Camera (ALFOSC) on the 2.5 m Nordic Optical Telescope (NOT) at the Roque de los Muchachos observatory (La Palma, Spain), the DEep Imaging Multi-Object Spectrograph (DEIMOS) on the 10 m Keck 2 telescope at the W. M. Keck Observatory (Hawaii, USA), and the Low Resolution ImagingSpectrometer (LRIS) on the 10 m Keck~1 telescope. The resolution of the spectra were measured using the [\ion{O}{III}] $\lambda 4959$ line (or, when resolved, the [\ion{S}{II}] $\lambda 6717, 6731$ lines) of the host galaxy. The epoch of the measurements are given relative to the time of SN discovery ($\rm JD~ 2456324.98$).}
\end{deluxetable}

\begin{deluxetable}{ccccc}
\tabletypesize{\scriptsize}
\tablewidth{0pt}
\tablecaption{Properties of iPTF13z light curve bumps\label{13z:bumps}}
\tablehead{
\colhead{Bump} &
\colhead{Maximum} &
\colhead{Duration} &
\colhead{Amplitude} &
\colhead{Band(s)} \\
\colhead{} &
\colhead{(days)} &
\colhead{(days)} &
\colhead{(mag)} &
\colhead{}}
\startdata
B$_1$ & 152 & 28 & 0.48 & $R$\\
B$_2$ & 372 & 35 & 0.75 & $g$,$R/r$,$i$\\
B$_3$ & 485 & $>$123 & 0.90 & $g$,$R/r$,$i$\\
B$_4$ & 570 & $>$71 & $\approx$0.40& $r$,$i$\\
B$_5$ & 732 & 20 & 0.44 & $g$,$R/r$,$i$
\enddata
\tablenotetext{*}{As bumps, we consider re-brightenings breaking the decline of the SN light curve, followed by renewed fading. The times of maxima, durations, and amplitudes given here are based on the linear fits to the decline-subtracted $r$-band light curve (Fig. \ref{fig:13zanalysis}b). Duration is measured from the time of positive crossing of $\Delta m = 0$ to the time of negative crossing. The `Band(s)' column specifies in which photometric band(s) the bumps were observed. The epochs of the maxima are given relative to time of SN discovery ($\rm JD~ 2456324.98$).}
\end{deluxetable}

\begin{deluxetable}{ccc}
\tabletypesize{\scriptsize}
\tablewidth{0pt}
\tablecaption{H$\alpha$ line velocities for iPTF13z\label{13z:havel}}
\tablehead{
\colhead{Epoch} &
\colhead{$FWHM~ \rm (H\alpha_b$)} &
\colhead{$FWHM~ \rm (H\alpha_n$)} \\
\colhead{(days)} &
\colhead{($\rm km ~ s^{-1}$)} &
\colhead{($\rm km ~ s^{-1}$)}}
\startdata
17.08 & 7968(869) & 262(45) \\ 
192.70 & 3407(173) & 469(11) \\ 
344.79 & 3757(249) & 569(12) \\ 
375.74 & 3471(241) & 726(57) \\ 
393.11 & 3209(529) & 135(19) \\ 
479.05 & 1648(106) & 149(12) \\ 
515.97 & 1486(71) & 131(12) \\ 
544.87 & 1215(165) & 286(16)
\enddata
\tablenotetext{*}{The values in parentheses are uncertainties calculated as the $1\sigma$ statistical error (from the least-squares fits to the H$\alpha$ line) and the systematic error added in quadrature. Systematic errors have been estimated by the difference between Gaussian and Lorentzian fits to the lines. The epoch of the measurements are given relative to the time of SN discovery ($\rm JD~ 2456324.98$).}
\end{deluxetable}

\begin{deluxetable}{cccc|c}
\tabletypesize{\scriptsize}
\tablewidth{0pt}
\tablecaption{H$\alpha$ line fluxes for iPTF13z\label{13z:haflux}}
\tablehead{
\colhead{Epoch} &
\colhead{$F~ \rm (H\alpha_b$)} &
\colhead{$F~ \rm (H\alpha_n$)} &
\colhead{$F~ \rm (H\alpha_{tot}$)} &
\colhead{$F~ \rm (H\alpha_{tot}^{num}$)} \\
\colhead{(days)} &
\colhead{} &
\colhead{} &
\colhead{} &
\colhead{}}
\startdata
17.08 & 4.56(0.11) & 0.31(0.11) & 5.30(0.24) & 4.24 \\
192.70 & 0.74(0.27) & 0.26(0.27) & 1.35(0.36) & 1.25 \\
344.79 & 1.68(0.28) & 0.45(0.28) & 2.31(0.39) & 2.13\\
375.74 & 1.26(0.44) & 0.46(0.44) & 2.09(0.60) & 1.95\\
393.11 & 0.70(0.06) & 0.09(0.06) & 0.97(0.10) & 0.92\\
479.05 & 0.80(0.15) & 0.17(0.15) & 1.19(0.20) & 1.23\\
515.97 & 0.57(0.11) & 0.11(0.11) & 0.84(0.14) & 0.89\\ 
544.87 & 0.19(0.12) & 0.14(0.12) & 0.48(0.17) & 0.52 
\enddata
\tablenotetext{*}{The line fluxes are given in units of $10^{-14}$ erg $\rm cm^{-2}$ $\rm s^{-1}$. The values in parentheses  are uncertainties calculated as the $1\sigma$ statistical error (from the least-squares fits to the H$\alpha$ line) and the systematic error added in quadrature. Systematic errors have been estimated by the difference between Gaussian and Lorentzian fits to the line. The epoch of the measurements are given relative to the time of SN discovery ($\rm JD~ 2456324.98$). For the narrow component, host-subtracted luminosities are given. To evaluate the total line flux found by the fitting procedure, numerical integration under the continuum subtracted spectra (in the wavelength range 6300$-$6800 \AA) is given in the right column.}
\end{deluxetable}

\end{document}